\documentclass[aps
,prl
,superscriptaddress
,twocolumn
]{revtex4-2}
\usepackage{graphicx} 
\usepackage{amsmath}
\usepackage[scaled]{helvet}
\usepackage[T1]{fontenc}
\usepackage{xcolor}
\usepackage{makecell}
\usepackage{bigints}
\usepackage{cancel}
\usepackage{bm}
\usepackage{xr-hyper}
\usepackage{hyperref}
\newcommand{\bra}[1]{\langle #1|}
\newcommand{\ket}[1]{|#1\rangle}
\newcommand{\braket}[2]{\langle #1|#2\rangle}

\def\bd{\textbf{d}}
\def\bA{\textbf{A}}

\def\bP{\textbf{P}}
\def\bR{\textbf{R}}

\def\dulR{{\underline{\underline{\bR}}}}

\def\ewf{\Phi_{\dulR}(t)}

\begin{document}

\title{Unifying Decoherence and Phase Evolution in Mixed Quantum–Classical Dynamics through Exact Factorization}

\date{\today}
\author{Jong-Kwon Ha}
\affiliation{Department of Chemistry, Dalhousie University, 6274 Coburg Rd, Halifax, NS B3H 4R2, Canada}
\author{Seong Ho Kim}
\affiliation{Department of Chemistry, Ulsan National Institute of Science and Technology (UNIST), 50 UNIST-gil, Ulju-gun, Ulsan 44919, South Korea}
\author{Seung Kyu Min}
\email{skmin@unist.ac.kr}
\affiliation{Department of Chemistry, Ulsan National Institute of Science and Technology (UNIST), 50 UNIST-gil, Ulju-gun, Ulsan 44919, South Korea}

\begin{abstract}
We propose mixed quantum–classical equations of motion that unify electronic coherence and phase evolution simultaneously within the exact factorization framework. Our derivation shows that incorporating the second-order electron–nuclear correlation terms from the exact coupled time-dependent Schrödinger equations is essential to recover both correct phase dynamics and complete electronic (de)coherence, including their effect on nuclear forces. Benchmark calculations on one- and two-dimensional model systems confirm that the approach accurately captures key nonadiabatic features. The equations therefore provide a rigorous first-principles foundation for mixed quantum–classical description of coupled electron–nuclear dynamics, bringing electronic coherence and phase evolution—long treated through separate heuristic corrections—into a single unified and systematically derived framework.
\end{abstract}

\maketitle
Understanding the coupled motion of electrons and nuclei in molecules remains one of the central challenges in theoretical chemistry and chemical physics. Fully quantum treatments of electron–nuclear dynamics are computationally prohibitive for systems of realistic size, owing to the scaling of the many-body molecular wavefunction. To overcome this limitation, mixed quantum–classical (MQC) methods~\cite{McLachlanMP1964,TullyJCP1990,CrespoCR2018} where the electrons are treated quantum mechanically while nuclei follow classical trajectories have become indispensable tools for simulating photochemical processes, charge and energy transfer, and ultrafast dynamics in molecules and materials~\cite{LinkerSciAdv2022,GopeSciAdv2022,DagarSciAdv2024,BorneNatChem2024,ChangNatPhys2025,LeeNatComm2025}. Among MQC methods, Ehrenfest dynamics~\cite{McLachlanMP1964} and trajectory-based surface hopping~\cite{TullyJCP1990} are the most widely applied. Despite their utility, they suffer from two fundamental shortcomings: the absence of a reliable mechanism for electronic decoherence, and the lack of a rigorous treatment of phase evolution between adiabatic states. 
Numerous correction schemes for decoherence and phase evolution have been proposed, often guided by semiclassical arguments, empirical adjustments, intuition from wave-packet models, or mapping of variables~\cite{ZhuJCP2004,GranucciJCP2007,XuJCP2019,ShaoJPCL2023,SubotnikJCP2011,ShenviJCP2011,MannouchJCP2023,WuJCTC2025}. Many approaches have successfully described electronic coherence or phase evolution, but they usually address only one of them or overlook their time-resolved nuclear wave-packet and electronic dynamics.

The exact factorization (XF) formalism provides a rigorous foundation for electron–nuclear dynamics by expressing the molecular wavefunction as a product of a nuclear wavefunction and a conditional electronic state~\cite{AbediPRL2010, AbediJCP2012}. This leads to coupled electronic and nuclear equations of motion that, in principle, include the full electron–nuclear correlation (ENC) effects. XF-based approaches have therefore attracted attention as a systematic framework for deriving improved MQC dynamics~\cite{MinPRL2015,AgostiniJCTC2016,GosselJCTC2018,TalottaJCTC2020,PieroniJCTC2021,ArribasJCP2023,GilJCP2024,HaJPCL2018,HaJCP2022,HanJCTC2023,DupuyJPCL2024,HanJCTC2025,HanJCTC2025_2}. In recent work, Arribas and Maitra showed that a projected quantum momentum (PQM) correction naturally arises from the XF formalism and plays a central role in describing electronic coherence~\cite{ArribasPRL2024}.
However, the correct phase-evolution effect in MQC dynamics has not been identified from the electron–nuclear correlation term within the XF framework. 
As a consequence, XF-based MQC methods fail to reproduce the proper relative phase evolution of the adiabatic states and, in turn, an important nonadiabatic feature, namely the St\"{u}ckelberg oscillation.

In this work, we derive new mixed quantum–classical equations of motion from the exact factorization formalism by including the terms from the previously neglected second-order ENC operator, revealing an additional phase-correction term that accompanies the PQM correction of comparable or smaller order in $\hbar$. This previously unidentified term governs the proper phase evolution of the Born–Oppenheimer (BO) coefficients. The combined inclusion of PQM and phase corrections provides a balanced and rigorous description of both decoherence and phase dynamics without relying on semiclassical approximations or heuristic arguments, offering deeper conceptual insight into ENC and a practical route to accurate molecular quantum dynamics. We validate the equations through benchmark calculations for the double-arch geometry (DAG)~\cite{SubotnikJCP2011}, dual avoided crossing (DAC)~\cite{TullyJCP1990},  and two-dimensional nonseparable (2DNS)~\cite{ShenviJCP2011} models, which demonstrate that both corrections are essential for reproducing characteristic nonadiabatic features such as St\"{u}ckelberg oscillations, intermediate-time electronic coherence, and the correct nuclear-density distributions.

The XF formalism expresses the molecular wavefunction $\ket{\Psi(\dulR, t)}$ as a single product of a time-dependent nuclear wavefunction $\chi(\dulR, t)$ and an electronic wavefunction $\ket{\Phi_\dulR(t)}$, $\ket{\Psi(\dulR, t)}=\chi(\dulR, t)\ket{\Phi_\dulR(t)}$ with partial normalization condition $\braket{\Phi_\dulR(t)}{\Phi_\dulR(t)}_e=1$ at any $\dulR$ and $t$, where $\dulR$ represents the nuclear degrees of freedom, and $\langle\cdots\rangle_e$ represents an integral over electronic degrees of freedom~\cite{AbediPRL2010, AbediJCP2012}.
Then, from the coupled time-dependent Schr\"odinger equations (TDSEs) for $\chi(\dulR,t)$ and $\ket{\Phi_\dulR(t)}$, the MQC equations can be derived~\cite{AgostiniJCTC2016, HanJCTC2023}.
The exact classical force for the nuclear trajectory $\dulR(t)$ is
\begin{equation}\label{eq:force}
\begin{split}
    \mathbf{F}_\nu = 
    \dot{\mathbf{P}}_\nu(t) = 
            -\nabla_\nu \tilde{\epsilon} + \dot{\bA}_\nu,
\end{split}
\end{equation}
and the electronic TDSE along the nuclear trajectory is
\begin{equation}\label{eq:tdse}
    i\hbar\dfrac{d}{dt}\ket{\Phi_\dulR(t)}=\left(\hat{H}_\mathrm{BO}+\hat{H}^{(1)}_\mathrm{ENC}+\hat{H}^{(2)}_\mathrm{ENC}-\tilde{\epsilon}\right)\ket{\Phi_\dulR(t)},
\end{equation}
where $\hat{H}_\mathrm{BO}$ is the BO Hamiltonian, and $\bA_\nu(\dulR,t) = \braket{\Phi_\dulR}{-i\hbar\nabla_\nu\Phi_\dulR}_e$ and $\tilde{\epsilon}(\dulR,t)=\bra{\Phi_\dulR(t)}(\hat{H}_\mathrm{BO}+\hat{H}^{(2)}_\mathrm{ENC}-i\hbar\tfrac{d}{dt})\ket{\Phi_\dulR(t)}_e$ are the time-dependent vector and scalar potentials, respectively.
Here, $\dot{f}$ represents $df/dt=\partial f/\partial t + \sum_\nu\dot{\bR}_\nu\cdot\nabla_\nu f$.
The first- and second-order electron–nuclear correlation (ENC) operators are
\begin{equation}
    \hat{H}^{(1)}_\mathrm{ENC}=-\sum_\nu\dfrac{\hbar^2}{2M_\nu}\bm{\mathcal{P}}_\nu\cdot\nabla_\nu\hat{\Gamma}_\dulR    
\end{equation}
and 
\begin{equation}
    \hat{H}^{(2)}_\mathrm{ENC}=-\sum_\nu\dfrac{\hbar^2}{2M_\nu}\left(\nabla_\nu^2\hat{\Gamma}_\dulR+\nabla_\nu\hat{\Gamma}_\dulR\cdot\nabla_\nu\hat{\Gamma}_\dulR\right),
\end{equation}
respectively, where $\bm{\mathcal{P}}_\nu=\nabla_\nu|\chi|^2/|\chi|^2$ is the total nuclear quantum momentum (QM), and $\hat{\Gamma}_\dulR=\ket{\Phi_\dulR}\bra{\Phi_\dulR}$ is the reduced electron density operator at a nuclear configuration $\dulR$. We note that all terms are evaluated along $\dulR(t)$ and the above equations are written in a general gauge without any specific choice of gauge.

Further expansion of eq~\ref{eq:tdse} in terms of BO basis states $\ket{\ewf}=\sum_j C_j\ket{\phi_j}=\sum_j |C_j|e^{iS_j/\hbar}\ket{\phi_j}$ satisfying $\hat{H}_\mathrm{BO}\ket{\phi_j}=\epsilon_j\ket{\phi_j}$ yields 
\begin{equation}\label{eq:cdot}
    \dot{C}_j=\dot{C}_j^\mathrm{Eh} + \dot{C}_j^\mathrm{QM} + \dot{C}_j^\mathrm{PQM} + \dot{C}_j^\mathrm{Div} + \dot{C}_j^\mathrm{Ph}.
\end{equation}
For the time evolution of BO coefficients, we have contributions from the Ehrenfest term $\dot{C}_j^\mathrm{Eh}=-\frac{i}{\hbar}\epsilon_jC_j-\sum_{\nu,k}\frac{\bP_\nu}{M_\nu}\cdot\bd_{\nu,jk}C_k$, the QM term $\dot{C}_j^\mathrm{QM} = -\sum_{\nu,k}\frac{\bm{\mathcal{P}}_\nu}{2M_\nu}\cdot\mathbf{D}_{\nu,jk}|C_k|^2C_j$ with $\mathbf{D}_{\nu,jk} = \nabla_\nu S_j - \nabla_\nu S_k$, the PQM~\cite{ArribasPRL2024} term 
\begin{equation}
    \dot{C}_j^\mathrm{PQM} = -\sum_{\nu,k}\frac{\bm{\mathcal{Q}}_{\nu,jk}}{2M_\nu}\cdot\mathbf{D}_{\nu,jk}|C_k|^2C_j    
\end{equation}
where $\bm{\mathcal{Q}}_{\nu,jk}=\nabla_\nu|C_j|^2/|C_j|^2+\nabla_\nu|C_k|^2/|C_k|^2$, the divergence term 
\begin{equation}
    \dot{C}_j^\mathrm{Div} = -\sum_{\nu,k}\frac{\nabla_\nu\cdot\mathbf{D}_{\nu,jk}}{2M_\nu}|C_k|^2C_j,
\end{equation}
and the phase correction term 
\begin{equation}
    \dot{C}_j^\mathrm{Ph} = \frac{i}{\hbar}(\tilde{\epsilon}-\sum_\nu\frac{|\nabla_\nu S_j-\bP_\nu|^2}{2M_\nu})C_j.
\end{equation}
Here we neglect the $O(\hbar^2)$ terms and the nonadiabatic coupling (NAC) terms ($\bd_{\nu,jk}=\braket{\phi_j}{\nabla_\nu\phi_k}_e$ and $g_{\nu,jk}=\braket{\phi_j}{\nabla^2_\nu\phi_k}_e$) in ENC terms. 

The nuclear force (eq~\ref{eq:force}) in the BO representation can be expressed as
\begin{equation}\label{eq:force_exp}
    \mathbf{F}_\nu = \mathbf{F}_\nu^\mathrm{Eh} + \mathbf{F}_\nu^\mathrm{QM} + \mathbf{F}_\nu^\mathrm{PQM} + \mathbf{F}_\nu^\mathrm{Div} + \mathbf{F}_\nu^\mathrm{Ph} + 
    \mathbf{F}_\nu^\mathrm{GI},
\end{equation}
where $\mathbf{F}_\nu^\mathrm{Eh}=-\sum_k|C_k|^2\nabla_\nu\epsilon_k - \sum_{k,l}C_kC_l^*(\epsilon_k-\epsilon_l)\bd_{\nu,lk}$, $\mathbf{F}_\nu^\mathrm{QM}=-\sum_{k,l}\mathbf{D}_{\nu,kl}\left(\sum_\mu\frac{\bm{\mathcal{P}}_{\mu}}{2M_\mu}\cdot\mathbf{D}_{\mu,kl}\right)|C_l|^2|C_k|^2$,
\begin{equation}
    \mathbf{F}_\nu^\mathrm{PQM}=-\sum_{k,l}\mathbf{D}_{\nu,kl}\left(\sum_\mu\dfrac{\bm{\mathcal{Q}}_{\mu,kl}}{2M_\mu}\cdot\mathbf{D}_{\mu,kl}\right)|C_l|^2|C_k|^2,
\end{equation}
\begin{equation}
    \mathbf{F}_\nu^\mathrm{Div}=-\sum_{k,l}\mathbf{D}_{\nu,kl}\left(\sum_\mu\dfrac{\nabla_\mu}{2M_\mu}\cdot\mathbf{D}_{\mu,kl}\right)|C_l|^2|C_k|^2,
\end{equation}
and
\begin{equation}\label{eq:force_gi_ph}
    \mathbf{F}_\nu^\mathrm{Ph} + \mathbf{F}_\nu^\mathrm{GI}=-\sum_{k,l}\sum_{\mu}\left(\dfrac{\mathbf{H}_{\nu\mu,kl}\mathbf{D}_{\mu,kl}}{2M_\mu}\right)|C_l|^2|C_k|^2,
\end{equation}
where $\mathbf{H}_{\nu\mu,kl}$ is a $3\times3$ cartesian sub-Hessian matrix of phase difference for nuclei $\nu$ and $\mu$, i.e. $\left[\mathbf{H}_{\nu\mu,kl}\right]_{pq} = \partial^2/\partial p_{\nu}\partial q_{\mu}(S_k-S_l)$ for cartesian components $p$ and $q$.
Each term directly corresponds to the contributions in eq~\ref{eq:cdot} except $\mathbf{F}_\nu^\mathrm{GI}$ from the gauge-invariant part of the time-dependent scalar potential, $\tilde{\epsilon}_\mathrm{GI} = \bra{\Phi_\dulR}\hat{H}^{(2)}_\mathrm{ENC}\ket{\Phi_\dulR}_e$ in eq~\ref{eq:force}, where a part of $\mathbf{F}_\nu^\mathrm{GI}$ cancels with $\mathbf{F}_\nu^\mathrm{Ph}$ to generate the net force in eq~\ref{eq:force_gi_ph}.
Detailed derivations are provided in Supporting Information.

The conventional XF-based MQC approaches, coupled-trajectory MQC (CTMQC)~\cite{MinPRL2015,AgostiniJCTC2016,GosselJCTC2018,TalottaJCTC2020,PieroniJCTC2021,ArribasJCP2023,GilJCP2024} or independent-trajectory approaches such as surface hopping dynamics based on XF (SHXF)~\cite{HaJPCL2018} and Ehrenfest dynamics based on XF (EhXF)~\cite{HaJCP2022}, only consider $\dot{C}_j = \dot{C}^\mathrm{Eh}_j+\dot{C}^\mathrm{QM}_j$ and $\mathbf{F}_\nu = \mathbf{F}_\nu^\mathrm{Eh} + \mathbf{F}_\nu^\mathrm{QM}$ obtained from the first-order ENC operator $\hat{H}_\mathrm{ENC}^{(1)}$ where $\dot{C}^\mathrm{QM}_j$ and $\mathbf{F}_\nu^\mathrm{QM}$ provide decoherence with nuclear wave-packet splitting. However, these approaches cannot provide accurate phase evolution~\cite{AgostiniJCTC2016,HaJCP2022}. Recently, Arribas and Maitra proposed that an additional correction to the total QM, $\dot{C}^\mathrm{PQM}_j$, is necessary to consider electronic coherence more accurately~\cite{ArribasPRL2024}. 

Here we claim that there are more additional terms to be considered at the same or less order in $\hbar$ as the QM correction, the phase term, $\dot{C}^\mathrm{Ph}_j$, and the divergence term, $\dot{C}^\mathrm{Div}_j$ which are crucial to describe correct BO phases and coherence simultaneously. These terms as well as $\dot{C}^\mathrm{PQM}_j$ are derived from the second ENC operator which has been previously overlooked, especially from the $\nabla_\nu^2\hat\Gamma_\dulR$ term in $\hat{H}^{(2)}_\mathrm{ENC}$.

The phase term, $\dot{C}^\mathrm{Ph}_j$, is of the same order as $-\frac{i}{\hbar}\epsilon_jC_j$ in $\dot{C}^\mathrm{Eh}_j$, which governs an additional phase evolution.
Here, $\tilde\epsilon$ is a global phase over the electronic wavefunction along a trajectory, and can therefore be neglected during time evolution, i.e. $\tilde{\epsilon}=0$, which is the actual gauge condition adopted in the numerical implementation and practical simulations.
Based on the semiclassical analysis, $\nabla_\nu S_i$ can be approximated to the momentum of a fictitious nuclear trajectory associated with state $i$. By assuming this momentum is parallel to the mean nuclear momentum $\bP_{\nu}$, we further approximate the state-wise momentum as $\bP_{\nu,i} = \alpha_i \bP_{\nu}$, where the scaling factor $\alpha_i$ is determined by the state-wise total energy ($E_\mathrm{tot}$) conservation condition:$\alpha_i = \sqrt{\max \left( 0, \frac{E_\mathrm{tot} - \epsilon_i}{\sum_\mu |\bP_\mu|^2 / 2M_\mu} \right)}$.
As a result, we obtain:
\begin{equation}
\begin{split}
    \dot{S}_j-\dot{S}_k
    =&-\sum_\nu\dfrac{\bP_{\nu,j}-\bP_{\nu,k}}{M_\nu}\cdot\bP_\nu
\end{split}
\end{equation}
from phases in $\dot{C}_j^\mathrm{Eh}$ and $\dot{C}_j^\mathrm{Ph}$.
This shows that the semiclassical phase correction that had previously been introduced in phase-corrected surface hopping (PCSH) heuristically~\cite{ShenviJCP2011} is actually an approximation of the phase term of XF.
In this context, we expect that the St\"{u}ckelberg oscillations can be captured with $\dot{C}^\mathrm{Ph}_j$ as they are in PCSH.

The divergence contribution, $\dot{C}^\mathrm{Div}_j$, contains the divergence of $\mathbf{D}_{\nu,jk}$, which requires the second derivatives of phases. 
Despite the difficulty to obtain the second derivatives, the $\dot{C}^\mathrm{Div}_j$ is essential for the conservation of BO population in the absence of NAC, i.e.
$\int d\dulR \dot{|C_j|^2}|\chi|^2 = \int \nabla_\nu\cdot(\sum_k|C_j|^2|C_k|^2|\chi|^2\mathbf{D}_{\nu,jk}) = 0$, which is obtained from $\dot{C}_j^\mathrm{QM}+\dot{C}_j^\mathrm{PQM}+\dot{C}_j^\mathrm{Div}$ since $\dot{C}_j^\mathrm{Eh}$ without NAC and $\dot{C}_j^\mathrm{Ph}$ does not contribute to the BO population transfer.
Therefore, in this work, we approximate the divergence term using constant pair-wise shift terms (See Supporting Information for the numerical implementation of the equations).

We benchmark the new equations of motion using the one-dimensional DAG~\cite{SubotnikJCP2011}, DAC~\cite{TullyJCP1990} and the 2DNS~\cite{ShenviJCP2011} models.
The model Hamiltonians and computational details are provided in the Supporting Information.

\begin{figure}[h!]
    \centering
    \includegraphics[width=1.0\linewidth]{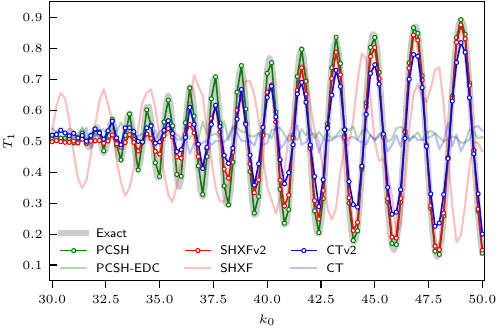}
    \caption{Transmission probability for the lower state ($T_1$) as a function of initial momentum $k_0$ for various dynamics methods. $T_1$ is obtained as $T_1 = \int_0^{\infty} dR \; |\chi_1(R,t_f)|^2$ for the exact wave-packet dynamics and $T_1 = \sum_{I \in \{R^{(I)}(t_f)>0\}} |C_1^{(I)}(t_f)|^2/N_\mathrm{traj}$ for the trajectory-based dynamics. $R^{(I)}(t)$ and $t_f$ denote the trajectory as a function of time $t$ and the final time of a simulation, respectively, with the total number of trajectories $N_\mathrm{traj}$.}
    \label{fig:RT_all}
\end{figure}
Figure~\ref{fig:RT_all} shows the transmission probability on the lower state, $T_1$, for the DAG model with various trajectory-based approaches in comparison with the exact wave-packet dynamics.
We refer to the results obtained without the additional terms as SHXF and CT, whereas those including the terms are denoted as SHXFv2 and CTv2, respectively.
The exact behavior shows an increase of oscillation amplitude as $k_0$ increases, from $\sim$ 0.5 at $k_0=30.0$ a.u.
As reported previously, PCSH can correctly reproduce the exact St\"{u}ckelberg oscillation as aimed~\cite{ShenviJCP2011}.
On the other hand, PCSH with the energy-based decoherence correction~\cite{GranucciJCP2007} (PCSH-EDC) removes the oscillation due to over-decoherence.
Similarly, CT, which only account for the decoherence, shows $T_1\sim0.5$ for all values of $k_0$.
Although SHXF shows an oscillatory behavior with respect to $k_0$, it deviates from the exact behavior.
CTv2 and SHXFv2 recover the correct St\"{u}ckelberg oscillation.
While CTv2 slightly underestimates the overall amplitude, SHXFv2 shows excellent agreement with the exact result at high $k_0$, and is qualitatively correct at low $k_0$.
This demonstrates that the additional phase correction term is essential for the correct time evolution of BO phases.
Based on the results in Figures~\ref{fig:pop_coh_35},~\ref{fig:ct2_spatial_k0_35}, and~\ref{fig:shxf2_spatial_k0_35}, we attribute the underestimation of the oscillation amplitude in the low-$k_{0}$ region to the numerical approximations rather than to the over-decoherence, although this needs to be further investigated in detail for more accurate simulations in the future.
Similarly, the DAC model, a standard Tully model which suffers from phase problems, shows that CTv2 and SHXFv2 well reproduce the correct T1 (Figure~\ref{fig:RT_dac}).

\begin{figure}
    \centering
    \includegraphics[width=1.0\linewidth]{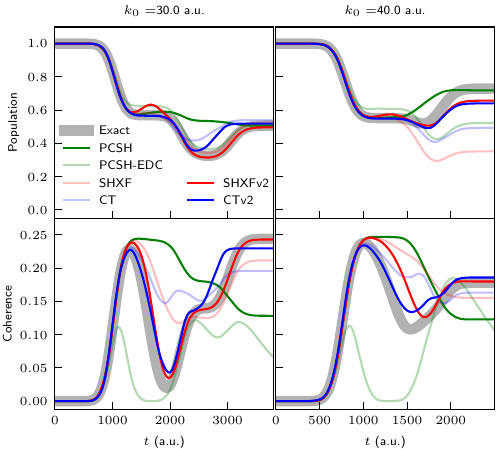}
    \caption{Time evolution of BO population $\langle|C_1|^2\rangle(t)$ (upper) and coherence $\langle|C_1C_2|^2\rangle(t)$ (lower) with different initial momenta $k_0 = 30.0$ and $40.0$ a.u.}
    \label{fig:pop_coh_all}
\end{figure}
Although reproducing the correct oscillation in $T_1$ is encouraging, it remains essential to evaluate the time propagation of wave packets to fully assess the accuracy.
Figure~\ref{fig:pop_coh_all} shows the time evolution of BO population for the lower state, $\langle|C_1|^2\rangle(t)$, and coherence, $\langle|C_1C_2|^2\rangle(t)$, with different initial momenta $k_0=30.0$ and 40.0 a.u. 
For exact wave-packet dynamics, $\langle O\rangle(t)$ is defined as $\langle O \rangle(t) = \int d\dulR \; O(\dulR,t)|\chi(\dulR,t)|^2$,
whereas $\langle O \rangle(t) \equiv \sum_{I}^{N_\mathrm{traj}} O^{(I)}(t)/N_\mathrm{traj}$ is used for MQC methods.
Since the DAG model contains two NAC regions, the initial population drop corresponds to passage through the first region, while the separated wave packets on upper and lower states afterward encounter the second region at different times.

PCSH reproduces the final BO populations well but fails to capture the intermediate dynamics for $k_0 = 30.0$ and shows overcoherence for all $k_0$ due to the absence of a decoherence correction. 
In contrast, PCSH-EDC exhibits over-decoherence, yielding final $\langle|C_1|^2\rangle$ to $\sim 0.5$ for all $k_0$.
All XF-based MQC methods reasonably reproduce the correct BO population dynamics for $k_0=30.0$ a.u. 
Notably, only CTv2 and SHXFv2 reproduce the accurate intermediate decoherence at $t \sim 2000$ a.u., highlighting the importance of the additional terms in CTv2 and SHXFv2.
For $k_0=40.0$ a.u., CT incorrectly predicts the final population to $\sim$~0.5, while SHXF yields the opposite direction of second population transfer.
Again, SHXFv2 and CTv2 successfully reproduce the correct BO population and coherence dynamics, including the population transfer at the second NAC region.
In these cases, CTv2 provides better coherence between the first and second NAC regions, whereas SHXFv2 exhibits slight overcoherence.

We also present the transmission probability $T_1$ and the time evolution of BO populations determined by the active states of the trajectories to test the internal consistency of SHXFv2, in Figures~\ref{fig:RT_run} and~\ref{fig:pop_run} of the Supporting Information, respectively. These results are in good agreement with those calculated from the electronic states of the trajectories (Figures~\ref{fig:RT_all} and~\ref{fig:pop_coh_all}), suggesting that both SHXF and SHXFv2 exhibit internal consistency. However, as previously noted, SHXF fails to accurately capture the transmission probability and BO population after passing through the second NAC region despite its internal consistency.

\begin{figure}
    \centering
    \includegraphics[width=1.0\linewidth]{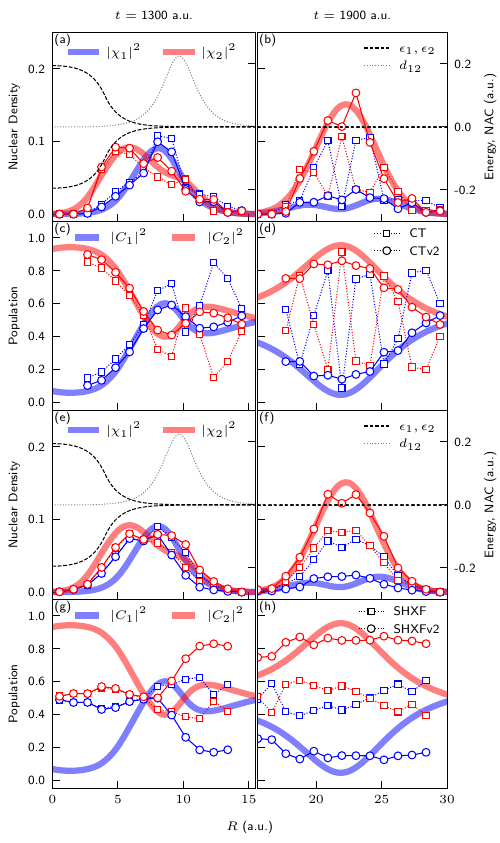}
    \caption{Spatial distribution of nuclear wave packets ($|\chi_i(R,t)|^2$) (a,b,e,f) and BO populations ($|C_i(R,t)|^2$) (c,d,g,h) at different time steps, $t=1300$ a.u. (a,c,e,g) and $1900$ a.u. (b,d,f,h) with the initial momentum $k_0=50.0$ a.u. for CT-based approaches (CT/CTv2) (a-d) and SHXF-based approaches (SHXF/SHXFv2) (e-h). The BO potential energies $\epsilon_{1/2}$ and the NAC $d_{12}$ are depicted to interpret wave-packet dynamics (a,b,e,f).}
    \label{fig:spatial_k0_50}
\end{figure}
In addition, to analyze the wave-packet dynamics, we construct the spatial distribution of the BO-projected time-dependent nuclear densities, $|\chi_i(R,t)|^2$, and BO populations, $|C_i(R,t)|^2$, for the DAG model. $|\chi_i(R,t)|^2$ are obtained from histograms of nuclear trajectories weighted by their corresponding BO populations, while $|C_i(R,t)|^2$ are calculated by $|C_i(R,t)|^2= |\chi_i(R,t)|^2/|\chi(R,t)|^2$ where the total nuclear density, $|\chi(R)|^2$, is the histograms of trajectories.
Figure~\ref{fig:spatial_k0_50} presents $|\chi_i(R,t)|^2$ and $|C_i(R,t)|^2$ from the XF-based approaches for large $k_0$ ($k_0=50.0$ a.u.), at $t=1300$ a.u. (arrival of nuclear wave packets at the second NAC region) and $t=1900$ a.u. (after passage through the second NAC region).
We observe a marked improvement in the spatial distributions with CTv2 and SHXFv2. CT exhibits oscillatory behavior arising from incorrect phase estimation in each BO state, while SHXF yields an overall inaccurate spatial distribution. By contrast, CTv2 reproduces almost identical results to the exact dynamics. SHXFv2 performs reasonably well at $t=1900$ a.u., but at $t=1300$ a.u. it produces strongly overlapping nuclear wave packets, likely due to a less accurate treatment of decoherence.

\begin{figure}
    \centering
    \includegraphics[width=1.0\linewidth]{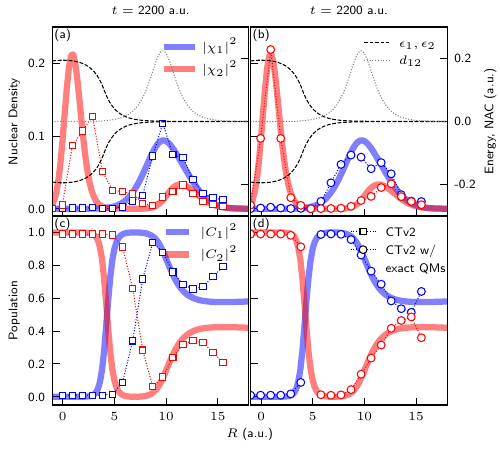}
    \caption{Spatial distribution of nuclear wave packets (a,b) and BO populations (c,d) at $t=2200$ a.u. with the initial momentum $k_0=30.0$ a.u. for CTv2 (a,c) and CTv2 with the QM $\mathbf{\mathcal{P}}_\nu$ and the PQM $\mathbf{\mathcal{Q}}_{\nu,kl}$ obtained from quantum dynamics (b,d). BO potentials and NAC are depicted as in Figure~\ref{fig:spatial_k0_50}.}
    \label{fig:ct2_spatial_k0_30}
\end{figure}
For $k_0=30.0$ a.u., the nuclear wave-packet splitting and the electronic decoherence effects are pronounced after passing the first NAC region due to a larger relative difference in the velocities of nuclear wave packets on different BO states, because of the small initial kinetic energy.
Thus, the well-separated nuclear wave packets pass the second NAC region sequentially.
Figure~\ref{fig:ct2_spatial_k0_30} shows nuclear wave packets and BO populations at $t=2200$ a.u. when $|\chi_1(R,t)|^2$ meets the second NAC region. 
CTv2 provides better wave-packet splitting and $|C_i(R,t)|^2$ compared to CT in Figure~\ref{fig:ct_spatial_k0_30}, while both deviate from the exact result.
On the other hand, the CTv2 with the exact QM and PQM recovers the correct $|\chi_i(R,t)|^2$ and $|C_i(R,t)|^2$, whereas CT (Figure~\ref{fig:ct_spatial_k0_30}) with the exact QM still deviates from the exact result.
Thus, the formulation is valid; the discrepancies mainly originate from the approximations to $\bm{\mathcal{P}}_\nu$ and $\bm{\mathcal{Q}}_{\nu,ik}$, as the true $|\chi_2(R,t)|^2$ becomes strongly non-Gaussian after the first NAC region for $k_0=30.0$ a.u.

The results from other approaches are provided in Figures~\ref{fig:ct_spatial_k0_30}–\ref{fig:pcsh_both_spatial_k0_50}. 
For all $k_0$, PCSH and PCSH-EDC yields either incorrect spatial decoherence or incorrect BO population after the second NAC region, respectively, whereas CTv2 and SHXFv2 exhibit clear enhancement compared to the other methods.

\begin{figure}[h!]
    \centering
    \includegraphics[width=1.0\linewidth]{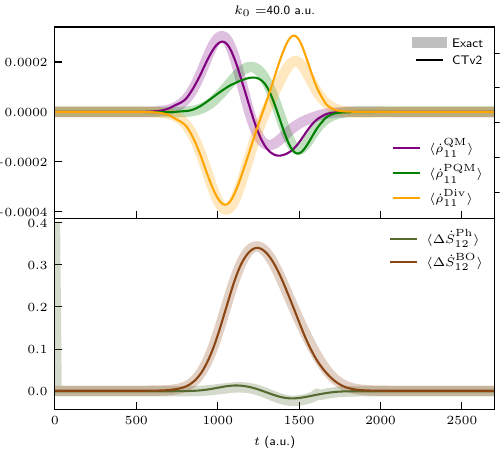}
    \caption{Time evolution of the individual components of the averaged time derivative of BO population for state 1 $\langle\dot{\rho}^\mathrm{QM/PQM/Div}_{11}\rangle(t)$ (upper) and the phase difference $\langle\Delta\dot{S}^\mathrm{Ph/BO}_{12}\rangle(t)$ (lower). Results obtained from CTv2 are shown in comparison with exact quantum wave packet dynamics for $k_0 = 40.0$~a.u.}
    \label{fig:element_evolution}
\end{figure}

Regarding the relative importance of the ENC terms, the electron-nuclear correlation effects can be separated into two groups governing the amplitude and phase evolutions of the BO coefficient, namely the decoherence ($\dot{C}^\mathrm{QM}_j$, $\dot{C}^\mathrm{PQM}_j$, and $\dot{C}^\mathrm{Div}_j$) and phase correction ($\dot{C}^\mathrm{Ph}_j$) terms, respectively. Also, considering the order of terms in $\hbar$, the phase correction would have much larger magnitude than the decoherence correction terms. Therefore, we analyze the decoherence and phase correction terms separately. For the phase correction term, we analyze the relative phase evolution. We show the time evolution of each component of $\langle\dot{\rho}_{11}\rangle$ and $\langle\Delta\dot{S}_{12}\rangle$ (Figure \ref{fig:element_evolution}), and the spatial characters of $F$, $\dot{\rho}_{11}$, and $\dot{S}_{12}$ at $t$=950 a.u. and 1250 a.u. (Figure \ref{fig:element_spatial}) for $k_0=40.0$, where $\dot{\rho}^{\mathrm{QM/PQM/Div}}_{11}=2\Re(\dot{C}^{\mathrm{QM/PQM/Div}}_1C^*_1)$ represents the population change induced by each ENC component and $\Delta \dot{S}^{\mathrm{Ph/BO}}_{12}=\dot{S}^{\mathrm{Ph/BO}}_1-\dot{S}^{\mathrm{Ph/BO}}_2$ is the relative phase change from $C^\mathrm{Ph}_j$ and the BO energy contribution in $C^\mathrm{Eh}_j$.

Overall, the CTv2 results well reproduce the exact results while the spatial shape of the divergence term contribution is averaged out across the trajectories due to the approximation in the numerical implementation. The PQM and divergence contributions have magnitudes comparable to those of the QM term, while their temporal evolution and spatial characteristics differ from the QM contribution. Meanwhile, the phase correction effect is much smaller in magnitude than the phase evolution effect from the BO energy: $\Delta\dot{S}^\mathrm{BO}_{12}=-(\epsilon_1-\epsilon_2)$. However, its effect is non-negligible as seen in Figure~\ref{fig:RT_all}, because the phase evolution is more sensitive than the magnitude of the BO coefficient. Therefore, these newly introduced terms are essential for capturing the correct electron-nuclear correlation effect on both decoherence and phase evolutions.

In addition, we investigate the independent implementation of the PQM or phase term in SHXF and CT in Figures~\ref{fig:RT_pqm_ph},~\ref{fig:pop_coh_pqm_ph},~\ref{fig:ct_pqm_ph_spatial_k0_40}, and~\ref{fig:shxf_pqm_ph_spatial_k0_40}. The results demonstrate that incorporating only the PQM term improves the description of intermediate decoherence but fails to capture the correct population dynamics when passing through the second NAC region. Conversely, including only the phase term yields accurate population dynamics at the second NAC region but cannot fully describe the intermediate decoherence. Thus, the full consideration is essential to provide correct dynamics.

\begin{figure}[h!]
    \centering
    \includegraphics[width=1.0\linewidth]{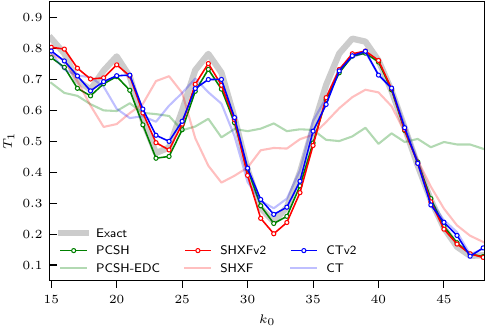}
    \caption{Transmission probability for the lower state ($T_1$) for the 2DNS model, as in Figure~\ref{fig:RT_all}.}
    \label{fig:RT_2dns_all}
\end{figure}
Finally, to test the additional corrections in higher-dimensional systems, we compute the $T_1$ for the 2DNS model.
As shown in Figure~\ref{fig:RT_2dns_all}, SHXFv2 and CTv2, along with PCSH, reproduce the correct oscillation pattern, whereas PCSH-EDC and SHXF fail, consistent with the DAG results.
CT works for $k_0>25$ a.u. but fails for $k_0=17$–25 a.u.
Thus, the phase and PQM corrections enable us to reproduce the quantum dynamics even in higher dimensions.

Regarding the numerical stability of CTv2 and SHXFv2, we investigate the mean absolute errors (MAEs) for norm conservation of the SHXFv2 and CTv2 simulations for $k_0=40.0$ a.u. for the DAG model. (Figure~\ref{fig:mae}) Both SHXFv2 and CTv2 exhibit reliable numerical stability comparable to that of SHXF and CT methods (MAE~$<10^{-8}$).  Although CTv2 and SHXFv2 involve additional terms from $\hat{H}^{(2)}_{ENC}$, each term from eqs~\ref{eq:cdot} and~\ref{eq:force_exp} can be obtained from the same ingredients required for CT and SHXF without any additional quantum chemical calculations. Consequently, the computational complexity of these new versions remains the same as the original methods.

In this work, we developed new mixed quantum–classical equations of motion for molecules within the exact factorization formalism. The derivation reveals that additional terms of the same or smaller order in $\hbar$ naturally arise from the full electron–nuclear correlation. In contrast to existing correction schemes, where decoherence and phase adjustments are often introduced heuristically and separately, the exact factorization framework provides both effects simultaneously and without ad hoc assumptions. The newly identified terms consist of the projected quantum momentum, which governs decoherence by correctly separating electronic populations, and a phase-correction term, which ensures accurate phase evolution of the nuclear wave packets.

Numerical benchmarks on representative one- and two-dimensional models demonstrate that these corrections are indispensable for reproducing key features of quantum dynamics, including St\"{u}ckelberg oscillations, intermediate-time (de)coherence, and the correct spatial distribution of nuclear wave packets. In particular, the enhanced performance of CTv2 and SHXFv2 underscores the importance of explicitly including both decoherence and phase corrections when simulating coupled electron–nuclear motion. Thus, our results establish that the exact factorization formalism provides not only a rigorous foundation for trajectory-based dynamics but also a systematic pathway for deriving corrections that capture the full scope of electron–nuclear correlation effects.

\begin{acknowledgments}
This research was supported by the National Research Foundation of Korea (NRF) funded by the Korean government (Ministry of Science and ICT (MSIT)) (RS-2023-NR119931, RS-2023-00257666, RS-2024-00455131).
\end{acknowledgments}
\bibliography{references}

\begin{thebibliography}{33}%
\makeatletter
\providecommand \@ifxundefined [1]{%
 \@ifx{#1\undefined}
}%
\providecommand \@ifnum [1]{%
 \ifnum #1\expandafter \@firstoftwo
 \else \expandafter \@secondoftwo
 \fi
}%
\providecommand \@ifx [1]{%
 \ifx #1\expandafter \@firstoftwo
 \else \expandafter \@secondoftwo
 \fi
}%
\providecommand \natexlab [1]{#1}%
\providecommand \enquote  [1]{``#1''}%
\providecommand \bibnamefont  [1]{#1}%
\providecommand \bibfnamefont [1]{#1}%
\providecommand \citenamefont [1]{#1}%
\providecommand \href@noop [0]{\@secondoftwo}%
\providecommand \href [0]{\begingroup \@sanitize@url \@href}%
\providecommand \@href[1]{\@@startlink{#1}\@@href}%
\providecommand \@@href[1]{\endgroup#1\@@endlink}%
\providecommand \@sanitize@url [0]{\catcode `\\12\catcode `\$12\catcode
  `\&12\catcode `\#12\catcode `\^12\catcode `\_12\catcode `\%12\relax}%
\providecommand \@@startlink[1]{}%
\providecommand \@@endlink[0]{}%
\providecommand \url  [0]{\begingroup\@sanitize@url \@url }%
\providecommand \@url [1]{\endgroup\@href {#1}{\urlprefix }}%
\providecommand \urlprefix  [0]{URL }%
\providecommand \Eprint [0]{\href }%
\providecommand \doibase [0]{https://doi.org/}%
\providecommand \selectlanguage [0]{\@gobble}%
\providecommand \bibinfo  [0]{\@secondoftwo}%
\providecommand \bibfield  [0]{\@secondoftwo}%
\providecommand \translation [1]{[#1]}%
\providecommand \BibitemOpen [0]{}%
\providecommand \bibitemStop [0]{}%
\providecommand \bibitemNoStop [0]{.\EOS\space}%
\providecommand \EOS [0]{\spacefactor3000\relax}%
\providecommand \BibitemShut  [1]{\csname bibitem#1\endcsname}%
\let\auto@bib@innerbib\@empty
\bibitem [{\citenamefont {McLachlan}(1964)}]{McLachlanMP1964}%
  \BibitemOpen
  \bibfield  {author} {\bibinfo {author} {\bibfnamefont {A.~D.}\ \bibnamefont
  {McLachlan}},\ }\bibfield  {title} {\bibinfo {title} {A variational solution
  of the time-dependent {S}chrodinger equation},\ }\href
  {https://doi.org/10.1080/00268976400100041} {\bibfield  {journal} {\bibinfo
  {journal} {Mol. Phys.}\ }\textbf {\bibinfo {volume} {8}},\ \bibinfo {pages}
  {39} (\bibinfo {year} {1964})}\BibitemShut {NoStop}%
\bibitem [{\citenamefont {Tully}(1990)}]{TullyJCP1990}%
  \BibitemOpen
  \bibfield  {author} {\bibinfo {author} {\bibfnamefont {J.~C.}\ \bibnamefont
  {Tully}},\ }\bibfield  {title} {\bibinfo {title} {Molecular dynamics with
  electronic transitions},\ }\href {https://doi.org/10.1063/1.459170}
  {\bibfield  {journal} {\bibinfo  {journal} {J. Chem. Phys.}\ }\textbf
  {\bibinfo {volume} {93}},\ \bibinfo {pages} {1061} (\bibinfo {year}
  {1990})}\BibitemShut {NoStop}%
\bibitem [{\citenamefont {Crespo-Otero}\ and\ \citenamefont
  {Barbatti}(2018)}]{CrespoCR2018}%
  \BibitemOpen
  \bibfield  {author} {\bibinfo {author} {\bibfnamefont {R.}~\bibnamefont
  {Crespo-Otero}}\ and\ \bibinfo {author} {\bibfnamefont {M.}~\bibnamefont
  {Barbatti}},\ }\bibfield  {title} {\bibinfo {title} {Recent advances and
  perspectives on nonadiabatic mixed quantum–classical dynamics},\ }\href
  {https://doi.org/10.1021/acs.chemrev.7b00577} {\bibfield  {journal} {\bibinfo
   {journal} {Chem. Rev.}\ }\textbf {\bibinfo {volume} {118}},\ \bibinfo
  {pages} {7026} (\bibinfo {year} {2018})}\BibitemShut {NoStop}%
\bibitem [{\citenamefont {Linker}\ \emph {et~al.}(2022)\citenamefont {Linker},
  \citenamefont {ichi Nomura}, \citenamefont {Aditya}, \citenamefont
  {Fukshima}, \citenamefont {Kalia}, \citenamefont {Krishnamoorthy},
  \citenamefont {Nakano}, \citenamefont {Rajak}, \citenamefont {Shimmura},
  \citenamefont {Shimojo},\ and\ \citenamefont {Vashishta}}]{LinkerSciAdv2022}%
  \BibitemOpen
  \bibfield  {author} {\bibinfo {author} {\bibfnamefont {T.}~\bibnamefont
  {Linker}}, \bibinfo {author} {\bibfnamefont {K.}~\bibnamefont {ichi Nomura}},
  \bibinfo {author} {\bibfnamefont {A.}~\bibnamefont {Aditya}}, \bibinfo
  {author} {\bibfnamefont {S.}~\bibnamefont {Fukshima}}, \bibinfo {author}
  {\bibfnamefont {R.~K.}\ \bibnamefont {Kalia}}, \bibinfo {author}
  {\bibfnamefont {A.}~\bibnamefont {Krishnamoorthy}}, \bibinfo {author}
  {\bibfnamefont {A.}~\bibnamefont {Nakano}}, \bibinfo {author} {\bibfnamefont
  {P.}~\bibnamefont {Rajak}}, \bibinfo {author} {\bibfnamefont
  {K.}~\bibnamefont {Shimmura}}, \bibinfo {author} {\bibfnamefont
  {F.}~\bibnamefont {Shimojo}},\ and\ \bibinfo {author} {\bibfnamefont
  {P.}~\bibnamefont {Vashishta}},\ }\bibfield  {title} {\bibinfo {title}
  {Exploring far-from-equilibrium ultrafast polarization control in
  ferroelectric oxides with excited-state neural network quantum molecular
  dynamics},\ }\href {https://doi.org/10.1126/sciadv.abk2625} {\bibfield
  {journal} {\bibinfo  {journal} {Sci. Adv.}\ }\textbf {\bibinfo {volume}
  {8}},\ \bibinfo {pages} {eabk2625} (\bibinfo {year} {2022})}\BibitemShut
  {NoStop}%
\bibitem [{\citenamefont {Gope}\ \emph {et~al.}(2022)\citenamefont {Gope},
  \citenamefont {Livshits}, \citenamefont {Bittner}, \citenamefont {Baer},\
  and\ \citenamefont {Strasser}}]{GopeSciAdv2022}%
  \BibitemOpen
  \bibfield  {author} {\bibinfo {author} {\bibfnamefont {K.}~\bibnamefont
  {Gope}}, \bibinfo {author} {\bibfnamefont {E.}~\bibnamefont {Livshits}},
  \bibinfo {author} {\bibfnamefont {D.~M.}\ \bibnamefont {Bittner}}, \bibinfo
  {author} {\bibfnamefont {R.}~\bibnamefont {Baer}},\ and\ \bibinfo {author}
  {\bibfnamefont {D.}~\bibnamefont {Strasser}},\ }\bibfield  {title} {\bibinfo
  {title} {An “inverse” harpoon mechanism},\ }\href
  {https://doi.org/10.1126/sciadv.abq8084} {\bibfield  {journal} {\bibinfo
  {journal} {Sci. Adv.}\ }\textbf {\bibinfo {volume} {8}},\ \bibinfo {pages}
  {eabq8084} (\bibinfo {year} {2022})}\BibitemShut {NoStop}%
\bibitem [{\citenamefont {Dagar}\ \emph {et~al.}(2024)\citenamefont {Dagar},
  \citenamefont {Zhang}, \citenamefont {Rosenberger}, \citenamefont {Linker},
  \citenamefont {Sousa-Castillo}, \citenamefont {Neuhaus}, \citenamefont
  {Mitra}, \citenamefont {Biswas}, \citenamefont {Feinberg}, \citenamefont
  {Summers}, \citenamefont {Nakano}, \citenamefont {Vashishta}, \citenamefont
  {Shimojo}, \citenamefont {Wu}, \citenamefont {Vera}, \citenamefont {Maier},
  \citenamefont {Cortés}, \citenamefont {Bergues},\ and\ \citenamefont
  {Kling}}]{DagarSciAdv2024}%
  \BibitemOpen
  \bibfield  {author} {\bibinfo {author} {\bibfnamefont {R.}~\bibnamefont
  {Dagar}}, \bibinfo {author} {\bibfnamefont {W.}~\bibnamefont {Zhang}},
  \bibinfo {author} {\bibfnamefont {P.}~\bibnamefont {Rosenberger}}, \bibinfo
  {author} {\bibfnamefont {T.~M.}\ \bibnamefont {Linker}}, \bibinfo {author}
  {\bibfnamefont {A.}~\bibnamefont {Sousa-Castillo}}, \bibinfo {author}
  {\bibfnamefont {M.}~\bibnamefont {Neuhaus}}, \bibinfo {author} {\bibfnamefont
  {S.}~\bibnamefont {Mitra}}, \bibinfo {author} {\bibfnamefont
  {S.}~\bibnamefont {Biswas}}, \bibinfo {author} {\bibfnamefont
  {A.}~\bibnamefont {Feinberg}}, \bibinfo {author} {\bibfnamefont {A.~M.}\
  \bibnamefont {Summers}}, \bibinfo {author} {\bibfnamefont {A.}~\bibnamefont
  {Nakano}}, \bibinfo {author} {\bibfnamefont {P.}~\bibnamefont {Vashishta}},
  \bibinfo {author} {\bibfnamefont {F.}~\bibnamefont {Shimojo}}, \bibinfo
  {author} {\bibfnamefont {J.}~\bibnamefont {Wu}}, \bibinfo {author}
  {\bibfnamefont {C.~C.}\ \bibnamefont {Vera}}, \bibinfo {author}
  {\bibfnamefont {S.~A.}\ \bibnamefont {Maier}}, \bibinfo {author}
  {\bibfnamefont {E.}~\bibnamefont {Cortés}}, \bibinfo {author} {\bibfnamefont
  {B.}~\bibnamefont {Bergues}},\ and\ \bibinfo {author} {\bibfnamefont {M.~F.}\
  \bibnamefont {Kling}},\ }\bibfield  {title} {\bibinfo {title} {Tracking
  surface charge dynamics on single nanoparticles},\ }\href
  {https://doi.org/10.1126/sciadv.adp1890} {\bibfield  {journal} {\bibinfo
  {journal} {Sci. Adv.}\ }\textbf {\bibinfo {volume} {10}},\ \bibinfo {pages}
  {eadp1890} (\bibinfo {year} {2024})}\BibitemShut {NoStop}%
\bibitem [{\citenamefont {{Borne}}\ \emph {et~al.}(2024)\citenamefont
  {{Borne}}, \citenamefont {{Cooper}}, \citenamefont {{Ashfold}}, \citenamefont
  {{Bachmann}}, \citenamefont {{Bhattacharyya}}, \citenamefont {{Boll}},
  \citenamefont {{Bonanomi}}, \citenamefont {{Bosch}}, \citenamefont
  {{Callegari}}, \citenamefont {{Centurion}}, \citenamefont {{Coreno}},
  \citenamefont {{Curchod}}, \citenamefont {{Danailov}}, \citenamefont
  {{Demidovich}}, \citenamefont {{Di Fraia}}, \citenamefont {{Erk}},
  \citenamefont {{Faccial{\`a}}}, \citenamefont {{Feifel}}, \citenamefont
  {{Forbes}}, \citenamefont {{Hansen}}, \citenamefont {{Holland}},
  \citenamefont {{Ingle}}, \citenamefont {{Lindh}}, \citenamefont {{Ma}},
  \citenamefont {{McGhee}}, \citenamefont {{Muvva}}, \citenamefont {{Nunes}},
  \citenamefont {{Odate}}, \citenamefont {{Pathak}}, \citenamefont {{Plekan}},
  \citenamefont {{Prince}}, \citenamefont {{Rebernik}}, \citenamefont
  {{Rouz{\'e}e}}, \citenamefont {{Rudenko}}, \citenamefont {{Simoncig}},
  \citenamefont {{Squibb}}, \citenamefont {{Venkatachalam}}, \citenamefont
  {{Vozzi}}, \citenamefont {{Weber}}, \citenamefont {{Kirrander}},\ and\
  \citenamefont {{Rolles}}}]{BorneNatChem2024}%
  \BibitemOpen
  \bibfield  {author} {\bibinfo {author} {\bibfnamefont {K.~D.}\ \bibnamefont
  {{Borne}}}, \bibinfo {author} {\bibfnamefont {J.~C.}\ \bibnamefont
  {{Cooper}}}, \bibinfo {author} {\bibfnamefont {M.~N.~R.}\ \bibnamefont
  {{Ashfold}}}, \bibinfo {author} {\bibfnamefont {J.}~\bibnamefont
  {{Bachmann}}}, \bibinfo {author} {\bibfnamefont {S.}~\bibnamefont
  {{Bhattacharyya}}}, \bibinfo {author} {\bibfnamefont {R.}~\bibnamefont
  {{Boll}}}, \bibinfo {author} {\bibfnamefont {M.}~\bibnamefont {{Bonanomi}}},
  \bibinfo {author} {\bibfnamefont {M.}~\bibnamefont {{Bosch}}}, \bibinfo
  {author} {\bibfnamefont {C.}~\bibnamefont {{Callegari}}}, \bibinfo {author}
  {\bibfnamefont {M.}~\bibnamefont {{Centurion}}}, \bibinfo {author}
  {\bibfnamefont {M.}~\bibnamefont {{Coreno}}}, \bibinfo {author}
  {\bibfnamefont {B.~F.~E.}\ \bibnamefont {{Curchod}}}, \bibinfo {author}
  {\bibfnamefont {M.~B.}\ \bibnamefont {{Danailov}}}, \bibinfo {author}
  {\bibfnamefont {A.}~\bibnamefont {{Demidovich}}}, \bibinfo {author}
  {\bibfnamefont {M.}~\bibnamefont {{Di Fraia}}}, \bibinfo {author}
  {\bibfnamefont {B.}~\bibnamefont {{Erk}}}, \bibinfo {author} {\bibfnamefont
  {D.}~\bibnamefont {{Faccial{\`a}}}}, \bibinfo {author} {\bibfnamefont
  {R.}~\bibnamefont {{Feifel}}}, \bibinfo {author} {\bibfnamefont {R.~J.~G.}\
  \bibnamefont {{Forbes}}}, \bibinfo {author} {\bibfnamefont {C.~S.}\
  \bibnamefont {{Hansen}}}, \bibinfo {author} {\bibfnamefont {D.~M.~P.}\
  \bibnamefont {{Holland}}}, \bibinfo {author} {\bibfnamefont {R.~A.}\
  \bibnamefont {{Ingle}}}, \bibinfo {author} {\bibfnamefont {R.}~\bibnamefont
  {{Lindh}}}, \bibinfo {author} {\bibfnamefont {L.}~\bibnamefont {{Ma}}},
  \bibinfo {author} {\bibfnamefont {H.~G.}\ \bibnamefont {{McGhee}}}, \bibinfo
  {author} {\bibfnamefont {S.~B.}\ \bibnamefont {{Muvva}}}, \bibinfo {author}
  {\bibfnamefont {J.~P.~F.}\ \bibnamefont {{Nunes}}}, \bibinfo {author}
  {\bibfnamefont {A.}~\bibnamefont {{Odate}}}, \bibinfo {author} {\bibfnamefont
  {S.}~\bibnamefont {{Pathak}}}, \bibinfo {author} {\bibfnamefont
  {O.}~\bibnamefont {{Plekan}}}, \bibinfo {author} {\bibfnamefont {K.~C.}\
  \bibnamefont {{Prince}}}, \bibinfo {author} {\bibfnamefont {P.}~\bibnamefont
  {{Rebernik}}}, \bibinfo {author} {\bibfnamefont {A.}~\bibnamefont
  {{Rouz{\'e}e}}}, \bibinfo {author} {\bibfnamefont {A.}~\bibnamefont
  {{Rudenko}}}, \bibinfo {author} {\bibfnamefont {A.}~\bibnamefont
  {{Simoncig}}}, \bibinfo {author} {\bibfnamefont {R.~J.}\ \bibnamefont
  {{Squibb}}}, \bibinfo {author} {\bibfnamefont {A.~S.}\ \bibnamefont
  {{Venkatachalam}}}, \bibinfo {author} {\bibfnamefont {C.}~\bibnamefont
  {{Vozzi}}}, \bibinfo {author} {\bibfnamefont {P.~M.}\ \bibnamefont
  {{Weber}}}, \bibinfo {author} {\bibfnamefont {A.}~\bibnamefont
  {{Kirrander}}},\ and\ \bibinfo {author} {\bibfnamefont {D.}~\bibnamefont
  {{Rolles}}},\ }\bibfield  {title} {\bibinfo {title} {{Ultrafast electronic
  relaxation pathways of the molecular photoswitch quadricyclane}},\ }\href
  {https://doi.org/10.1038/s41557-023-01420-w} {\bibfield  {journal} {\bibinfo
  {journal} {Nat. Chem.}\ }\textbf {\bibinfo {volume} {16}},\ \bibinfo {pages}
  {499} (\bibinfo {year} {2024})}\BibitemShut {NoStop}%
\bibitem [{\citenamefont {{Chang}}\ \emph {et~al.}(2025)\citenamefont
  {{Chang}}, \citenamefont {{Balciunas}}, \citenamefont {{Yin}}, \citenamefont
  {{Sapunar}}, \citenamefont {{Tenorio}}, \citenamefont {{Paul}}, \citenamefont
  {{Tsuru}}, \citenamefont {{Koch}}, \citenamefont {{Wolf}}, \citenamefont
  {{Coriani}},\ and\ \citenamefont {{W{\"o}rner}}}]{ChangNatPhys2025}%
  \BibitemOpen
  \bibfield  {author} {\bibinfo {author} {\bibfnamefont {Y.-P.}\ \bibnamefont
  {{Chang}}}, \bibinfo {author} {\bibfnamefont {T.}~\bibnamefont
  {{Balciunas}}}, \bibinfo {author} {\bibfnamefont {Z.}~\bibnamefont {{Yin}}},
  \bibinfo {author} {\bibfnamefont {M.}~\bibnamefont {{Sapunar}}}, \bibinfo
  {author} {\bibfnamefont {B.~N.~C.}\ \bibnamefont {{Tenorio}}}, \bibinfo
  {author} {\bibfnamefont {A.~C.}\ \bibnamefont {{Paul}}}, \bibinfo {author}
  {\bibfnamefont {S.}~\bibnamefont {{Tsuru}}}, \bibinfo {author} {\bibfnamefont
  {H.}~\bibnamefont {{Koch}}}, \bibinfo {author} {\bibfnamefont {J.-P.}\
  \bibnamefont {{Wolf}}}, \bibinfo {author} {\bibfnamefont {S.}~\bibnamefont
  {{Coriani}}},\ and\ \bibinfo {author} {\bibfnamefont {H.~J.}\ \bibnamefont
  {{W{\"o}rner}}},\ }\bibfield  {title} {\bibinfo {title} {{Electronic dynamics
  created at conical intersections and its dephasing in aqueous solution}},\
  }\href {https://doi.org/10.1038/s41567-024-02703-w} {\bibfield  {journal}
  {\bibinfo  {journal} {Nat. Phys.}\ }\textbf {\bibinfo {volume} {21}},\
  \bibinfo {pages} {137} (\bibinfo {year} {2025})}\BibitemShut {NoStop}%
\bibitem [{\citenamefont {{Lee}}\ \emph {et~al.}(2025)\citenamefont {{Lee}},
  \citenamefont {{Filatov}},\ and\ \citenamefont {{Min}}}]{LeeNatComm2025}%
  \BibitemOpen
  \bibfield  {author} {\bibinfo {author} {\bibfnamefont {I.~S.}\ \bibnamefont
  {{Lee}}}, \bibinfo {author} {\bibfnamefont {M.}~\bibnamefont {{Filatov}}},\
  and\ \bibinfo {author} {\bibfnamefont {S.~K.}\ \bibnamefont {{Min}}},\
  }\bibfield  {title} {\bibinfo {title} {{Dynamics of a light-driven molecular
  rotary motor in an optical cavity}},\ }\href
  {https://doi.org/10.1038/s41467-025-59607-3} {\bibfield  {journal} {\bibinfo
  {journal} {Nat. Commun.}\ }\textbf {\bibinfo {volume} {16}},\ \bibinfo {eid}
  {4554} (\bibinfo {year} {2025})}\BibitemShut {NoStop}%
\bibitem [{\citenamefont {Zhu}\ \emph {et~al.}(2004)\citenamefont {Zhu},
  \citenamefont {Nangia}, \citenamefont {Jasper},\ and\ \citenamefont
  {Truhlar}}]{ZhuJCP2004}%
  \BibitemOpen
  \bibfield  {author} {\bibinfo {author} {\bibfnamefont {C.}~\bibnamefont
  {Zhu}}, \bibinfo {author} {\bibfnamefont {S.}~\bibnamefont {Nangia}},
  \bibinfo {author} {\bibfnamefont {A.~W.}\ \bibnamefont {Jasper}},\ and\
  \bibinfo {author} {\bibfnamefont {D.~G.}\ \bibnamefont {Truhlar}},\
  }\bibfield  {title} {\bibinfo {title} {{Coherent switching with decay of
  mixing: An improved treatment of electronic coherence for
  non-Born–Oppenheimer trajectories}},\ }\href
  {https://doi.org/10.1063/1.1793991} {\bibfield  {journal} {\bibinfo
  {journal} {J. Chem. Phys.}\ }\textbf {\bibinfo {volume} {121}},\ \bibinfo
  {pages} {7658} (\bibinfo {year} {2004})}\BibitemShut {NoStop}%
\bibitem [{\citenamefont {Granucci}\ and\ \citenamefont
  {Persico}(2007)}]{GranucciJCP2007}%
  \BibitemOpen
  \bibfield  {author} {\bibinfo {author} {\bibfnamefont {G.}~\bibnamefont
  {Granucci}}\ and\ \bibinfo {author} {\bibfnamefont {M.}~\bibnamefont
  {Persico}},\ }\bibfield  {title} {\bibinfo {title} {Critical appraisal of the
  fewest switches algorithm for surface hopping},\ }\href
  {https://doi.org/10.1063/1.2715585} {\bibfield  {journal} {\bibinfo
  {journal} {J. Chem. Phys.}\ }\textbf {\bibinfo {volume} {126}},\ \bibinfo
  {pages} {134114} (\bibinfo {year} {2007})}\BibitemShut {NoStop}%
\bibitem [{\citenamefont {Xu}\ and\ \citenamefont {Wang}(2019)}]{XuJCP2019}%
  \BibitemOpen
  \bibfield  {author} {\bibinfo {author} {\bibfnamefont {J.}~\bibnamefont
  {Xu}}\ and\ \bibinfo {author} {\bibfnamefont {L.}~\bibnamefont {Wang}},\
  }\bibfield  {title} {\bibinfo {title} {{Branching corrected surface hopping:
  Resetting wavefunction coefficients based on judgement of wave packet
  reflection}},\ }\href {https://doi.org/10.1063/1.5090927} {\bibfield
  {journal} {\bibinfo  {journal} {J. Chem. Phys.}\ }\textbf {\bibinfo {volume}
  {150}},\ \bibinfo {pages} {164101} (\bibinfo {year} {2019})}\BibitemShut
  {NoStop}%
\bibitem [{\citenamefont {Shao}\ \emph {et~al.}(2023)\citenamefont {Shao},
  \citenamefont {Shi}, \citenamefont {Xu},\ and\ \citenamefont
  {Wang}}]{ShaoJPCL2023}%
  \BibitemOpen
  \bibfield  {author} {\bibinfo {author} {\bibfnamefont {C.}~\bibnamefont
  {Shao}}, \bibinfo {author} {\bibfnamefont {Z.}~\bibnamefont {Shi}}, \bibinfo
  {author} {\bibfnamefont {J.}~\bibnamefont {Xu}},\ and\ \bibinfo {author}
  {\bibfnamefont {L.}~\bibnamefont {Wang}},\ }\bibfield  {title} {\bibinfo
  {title} {Learning decoherence time formulas for surface hopping from quantum
  dynamics},\ }\href {https://doi.org/10.1021/acs.jpclett.3c02019} {\bibfield
  {journal} {\bibinfo  {journal} {J. Phys. Chem. Lett.}\ }\textbf {\bibinfo
  {volume} {14}},\ \bibinfo {pages} {7680} (\bibinfo {year}
  {2023})}\BibitemShut {NoStop}%
\bibitem [{\citenamefont {Subotnik}\ and\ \citenamefont
  {Shenvi}(2011)}]{SubotnikJCP2011}%
  \BibitemOpen
  \bibfield  {author} {\bibinfo {author} {\bibfnamefont {J.~E.}\ \bibnamefont
  {Subotnik}}\ and\ \bibinfo {author} {\bibfnamefont {N.}~\bibnamefont
  {Shenvi}},\ }\bibfield  {title} {\bibinfo {title} {A new approach to
  decoherence and momentum rescaling in the surface hopping algorithm},\ }\href
  {https://doi.org/10.1063/1.3506779} {\bibfield  {journal} {\bibinfo
  {journal} {J. Chem. Phys.}\ }\textbf {\bibinfo {volume} {134}},\ \bibinfo
  {pages} {024105} (\bibinfo {year} {2011})}\BibitemShut {NoStop}%
\bibitem [{\citenamefont {Shenvi}\ \emph {et~al.}(2011)\citenamefont {Shenvi},
  \citenamefont {Subotnik},\ and\ \citenamefont {Yang}}]{ShenviJCP2011}%
  \BibitemOpen
  \bibfield  {author} {\bibinfo {author} {\bibfnamefont {N.}~\bibnamefont
  {Shenvi}}, \bibinfo {author} {\bibfnamefont {J.~E.}\ \bibnamefont
  {Subotnik}},\ and\ \bibinfo {author} {\bibfnamefont {W.}~\bibnamefont
  {Yang}},\ }\bibfield  {title} {\bibinfo {title} {{Phase-corrected surface
  hopping: Correcting the phase evolution of the electronic wavefunction}},\
  }\href {https://doi.org/10.1063/1.3603447} {\bibfield  {journal} {\bibinfo
  {journal} {J. Chem. Phys.}\ }\textbf {\bibinfo {volume} {135}},\ \bibinfo
  {pages} {024101} (\bibinfo {year} {2011})}\BibitemShut {NoStop}%
\bibitem [{\citenamefont {Mannouch}\ and\ \citenamefont
  {Richardson}(2023)}]{MannouchJCP2023}%
  \BibitemOpen
  \bibfield  {author} {\bibinfo {author} {\bibfnamefont {J.~R.}\ \bibnamefont
  {Mannouch}}\ and\ \bibinfo {author} {\bibfnamefont {J.~O.}\ \bibnamefont
  {Richardson}},\ }\bibfield  {title} {\bibinfo {title} {A mapping approach to
  surface hopping},\ }\href {https://doi.org/10.1063/5.0139734} {\bibfield
  {journal} {\bibinfo  {journal} {J. Chem. Phys.}\ }\textbf {\bibinfo {volume}
  {158}},\ \bibinfo {pages} {104111} (\bibinfo {year} {2023})}\BibitemShut
  {NoStop}%
\bibitem [{\citenamefont {Wu}\ \emph {et~al.}(2025)\citenamefont {Wu},
  \citenamefont {Li}, \citenamefont {He}, \citenamefont {Cheng}, \citenamefont
  {Ren},\ and\ \citenamefont {Liu}}]{WuJCTC2025}%
  \BibitemOpen
  \bibfield  {author} {\bibinfo {author} {\bibfnamefont {B.}~\bibnamefont
  {Wu}}, \bibinfo {author} {\bibfnamefont {B.}~\bibnamefont {Li}}, \bibinfo
  {author} {\bibfnamefont {X.}~\bibnamefont {He}}, \bibinfo {author}
  {\bibfnamefont {X.}~\bibnamefont {Cheng}}, \bibinfo {author} {\bibfnamefont
  {J.}~\bibnamefont {Ren}},\ and\ \bibinfo {author} {\bibfnamefont
  {J.}~\bibnamefont {Liu}},\ }\bibfield  {title} {\bibinfo {title}
  {Nonadiabatic field: A conceptually novel approach for nonadiabatic quantum
  molecular dynamics},\ }\href {https://doi.org/10.1021/acs.jctc.5c00181}
  {\bibfield  {journal} {\bibinfo  {journal} {J. Chem. Theory Comput.}\
  }\textbf {\bibinfo {volume} {21}},\ \bibinfo {pages} {3775} (\bibinfo {year}
  {2025})}\BibitemShut {NoStop}%
\bibitem [{\citenamefont {Abedi}\ \emph {et~al.}(2010)\citenamefont {Abedi},
  \citenamefont {Maitra},\ and\ \citenamefont {Gross}}]{AbediPRL2010}%
  \BibitemOpen
  \bibfield  {author} {\bibinfo {author} {\bibfnamefont {A.}~\bibnamefont
  {Abedi}}, \bibinfo {author} {\bibfnamefont {N.~T.}\ \bibnamefont {Maitra}},\
  and\ \bibinfo {author} {\bibfnamefont {E.~K.~U.}\ \bibnamefont {Gross}},\
  }\bibfield  {title} {\bibinfo {title} {{Exact Factorization of the
  Time-Dependent Electron-Nuclear Wave Function}},\ }\href
  {https://doi.org/10.1103/PhysRevLett.105.123002} {\bibfield  {journal}
  {\bibinfo  {journal} {Phys. Rev. Lett.}\ }\textbf {\bibinfo {volume} {105}},\
  \bibinfo {pages} {123002} (\bibinfo {year} {2010})}\BibitemShut {NoStop}%
\bibitem [{\citenamefont {Abedi}\ \emph {et~al.}(2012)\citenamefont {Abedi},
  \citenamefont {Maitra},\ and\ \citenamefont {Gross}}]{AbediJCP2012}%
  \BibitemOpen
  \bibfield  {author} {\bibinfo {author} {\bibfnamefont {A.}~\bibnamefont
  {Abedi}}, \bibinfo {author} {\bibfnamefont {N.~T.}\ \bibnamefont {Maitra}},\
  and\ \bibinfo {author} {\bibfnamefont {E.~K.~U.}\ \bibnamefont {Gross}},\
  }\bibfield  {title} {\bibinfo {title} {{Correlated electron-nuclear dynamics:
  Exact factorization of the molecular wavefunction}},\ }\href
  {https://doi.org/10.1063/1.4745836} {\bibfield  {journal} {\bibinfo
  {journal} {J. Chem. Phys.}\ }\textbf {\bibinfo {volume} {137}},\ \bibinfo
  {pages} {22A530} (\bibinfo {year} {2012})}\BibitemShut {NoStop}%
\bibitem [{\citenamefont {Min}\ \emph {et~al.}(2015)\citenamefont {Min},
  \citenamefont {Agostini},\ and\ \citenamefont {Gross}}]{MinPRL2015}%
  \BibitemOpen
  \bibfield  {author} {\bibinfo {author} {\bibfnamefont {S.~K.}\ \bibnamefont
  {Min}}, \bibinfo {author} {\bibfnamefont {F.}~\bibnamefont {Agostini}},\ and\
  \bibinfo {author} {\bibfnamefont {E.~K.~U.}\ \bibnamefont {Gross}},\
  }\bibfield  {title} {\bibinfo {title} {{Coupled-Trajectory Quantum-Classical
  Approach to Electronic Decoherence in Nonadiabatic Processes}},\ }\href
  {https://doi.org/10.1103/PhysRevLett.115.073001} {\bibfield  {journal}
  {\bibinfo  {journal} {Phys. Rev. Lett.}\ }\textbf {\bibinfo {volume} {115}},\
  \bibinfo {pages} {073001} (\bibinfo {year} {2015})}\BibitemShut {NoStop}%
\bibitem [{\citenamefont {Agostini}\ \emph {et~al.}(2016)\citenamefont
  {Agostini}, \citenamefont {Min}, \citenamefont {Abedi},\ and\ \citenamefont
  {Gross}}]{AgostiniJCTC2016}%
  \BibitemOpen
  \bibfield  {author} {\bibinfo {author} {\bibfnamefont {F.}~\bibnamefont
  {Agostini}}, \bibinfo {author} {\bibfnamefont {S.~K.}\ \bibnamefont {Min}},
  \bibinfo {author} {\bibfnamefont {A.}~\bibnamefont {Abedi}},\ and\ \bibinfo
  {author} {\bibfnamefont {E.~K.~U.}\ \bibnamefont {Gross}},\ }\bibfield
  {title} {\bibinfo {title} {{Quantum-Classical Nonadiabatic Dynamics: Coupled-
  vs Independent-Trajectory Methods}},\ }\href
  {https://doi.org/10.1021/acs.jctc.5b01180} {\bibfield  {journal} {\bibinfo
  {journal} {J. Chem. Theory Comput.}\ }\textbf {\bibinfo {volume} {12}},\
  \bibinfo {pages} {2127} (\bibinfo {year} {2016})}\BibitemShut {NoStop}%
\bibitem [{\citenamefont {Gossel}\ \emph {et~al.}(2018)\citenamefont {Gossel},
  \citenamefont {Agostini},\ and\ \citenamefont {Maitra}}]{GosselJCTC2018}%
  \BibitemOpen
  \bibfield  {author} {\bibinfo {author} {\bibfnamefont {G.~H.}\ \bibnamefont
  {Gossel}}, \bibinfo {author} {\bibfnamefont {F.}~\bibnamefont {Agostini}},\
  and\ \bibinfo {author} {\bibfnamefont {N.~T.}\ \bibnamefont {Maitra}},\
  }\bibfield  {title} {\bibinfo {title} {{Coupled-Trajectory Mixed
  Quantum-Classical Algorithm: A Deconstruction}},\ }\href
  {https://doi.org/10.1021/acs.jctc.8b00449} {\bibfield  {journal} {\bibinfo
  {journal} {J. Chem. Theory Comput.}\ }\textbf {\bibinfo {volume} {14}},\
  \bibinfo {pages} {4513} (\bibinfo {year} {2018})}\BibitemShut {NoStop}%
\bibitem [{\citenamefont {Talotta}\ \emph {et~al.}(2020)\citenamefont
  {Talotta}, \citenamefont {Morisset}, \citenamefont {Rougeau}, \citenamefont
  {Lauvergnat},\ and\ \citenamefont {Agostini}}]{TalottaJCTC2020}%
  \BibitemOpen
  \bibfield  {author} {\bibinfo {author} {\bibfnamefont {F.}~\bibnamefont
  {Talotta}}, \bibinfo {author} {\bibfnamefont {S.}~\bibnamefont {Morisset}},
  \bibinfo {author} {\bibfnamefont {N.}~\bibnamefont {Rougeau}}, \bibinfo
  {author} {\bibfnamefont {D.}~\bibnamefont {Lauvergnat}},\ and\ \bibinfo
  {author} {\bibfnamefont {F.}~\bibnamefont {Agostini}},\ }\bibfield  {title}
  {\bibinfo {title} {{Internal Conversion and Intersystem Crossing with the
  Exact Factorization}},\ }\href {https://doi.org/10.1021/acs.jctc.0c00493}
  {\bibfield  {journal} {\bibinfo  {journal} {J. Chem. Theory Comput.}\
  }\textbf {\bibinfo {volume} {16}},\ \bibinfo {pages} {4833} (\bibinfo {year}
  {2020})}\BibitemShut {NoStop}%
\bibitem [{\citenamefont {Pieroni}\ and\ \citenamefont
  {Agostini}(2021)}]{PieroniJCTC2021}%
  \BibitemOpen
  \bibfield  {author} {\bibinfo {author} {\bibfnamefont {C.}~\bibnamefont
  {Pieroni}}\ and\ \bibinfo {author} {\bibfnamefont {F.}~\bibnamefont
  {Agostini}},\ }\bibfield  {title} {\bibinfo {title} {{Nonadiabatic Dynamics
  with Coupled Trajectories}},\ }\href
  {https://doi.org/10.1021/acs.jctc.1c00438} {\bibfield  {journal} {\bibinfo
  {journal} {J. Chem. Theory Comput.}\ }\textbf {\bibinfo {volume} {17}},\
  \bibinfo {pages} {5969} (\bibinfo {year} {2021})}\BibitemShut {NoStop}%
\bibitem [{\citenamefont {Arribas}\ and\ \citenamefont
  {Maitra}(2023)}]{ArribasJCP2023}%
  \BibitemOpen
  \bibfield  {author} {\bibinfo {author} {\bibfnamefont {E.~V.}\ \bibnamefont
  {Arribas}}\ and\ \bibinfo {author} {\bibfnamefont {N.~T.}\ \bibnamefont
  {Maitra}},\ }\bibfield  {title} {\bibinfo {title} {Energy-conserving coupled
  trajectory mixed quantum–classical dynamics},\ }\href
  {https://doi.org/10.1063/5.0149116} {\bibfield  {journal} {\bibinfo
  {journal} {J. Chem. Phys.}\ }\textbf {\bibinfo {volume} {158}},\ \bibinfo
  {pages} {161105} (\bibinfo {year} {2023})}\BibitemShut {NoStop}%
\bibitem [{\citenamefont {Sangiogo~Gil}\ \emph {et~al.}(2024)\citenamefont
  {Sangiogo~Gil}, \citenamefont {Lauvergnat},\ and\ \citenamefont
  {Agostini}}]{GilJCP2024}%
  \BibitemOpen
  \bibfield  {author} {\bibinfo {author} {\bibfnamefont {E.}~\bibnamefont
  {Sangiogo~Gil}}, \bibinfo {author} {\bibfnamefont {D.}~\bibnamefont
  {Lauvergnat}},\ and\ \bibinfo {author} {\bibfnamefont {F.}~\bibnamefont
  {Agostini}},\ }\bibfield  {title} {\bibinfo {title} {Exact factorization of
  the photon–electron–nuclear wavefunction: Formulation and
  coupled-trajectory dynamics},\ }\href {https://doi.org/10.1063/5.0224779}
  {\bibfield  {journal} {\bibinfo  {journal} {J. Chem. Phys.}\ }\textbf
  {\bibinfo {volume} {161}},\ \bibinfo {pages} {084112} (\bibinfo {year}
  {2024})}\BibitemShut {NoStop}%
\bibitem [{\citenamefont {Ha}\ \emph {et~al.}(2018)\citenamefont {Ha},
  \citenamefont {Lee},\ and\ \citenamefont {Min}}]{HaJPCL2018}%
  \BibitemOpen
  \bibfield  {author} {\bibinfo {author} {\bibfnamefont {J.-K.}\ \bibnamefont
  {Ha}}, \bibinfo {author} {\bibfnamefont {I.~S.}\ \bibnamefont {Lee}},\ and\
  \bibinfo {author} {\bibfnamefont {S.~K.}\ \bibnamefont {Min}},\ }\bibfield
  {title} {\bibinfo {title} {{Surface Hopping Dynamics beyond Nonadiabatic
  Couplings for Quantum Coherence}},\ }\href
  {https://doi.org/10.1021/acs.jpclett.8b00060} {\bibfield  {journal} {\bibinfo
   {journal} {J. Phys. Chem. Lett.}\ }\textbf {\bibinfo {volume} {9}},\
  \bibinfo {pages} {1097} (\bibinfo {year} {2018})}\BibitemShut {NoStop}%
\bibitem [{\citenamefont {Ha}\ and\ \citenamefont {Min}(2022)}]{HaJCP2022}%
  \BibitemOpen
  \bibfield  {author} {\bibinfo {author} {\bibfnamefont {J.-K.}\ \bibnamefont
  {Ha}}\ and\ \bibinfo {author} {\bibfnamefont {S.~K.}\ \bibnamefont {Min}},\
  }\bibfield  {title} {\bibinfo {title} {Independent trajectory mixed
  quantum-classical approaches based on the exact factorization},\ }\href
  {https://doi.org/10.1063/5.0084493} {\bibfield  {journal} {\bibinfo
  {journal} {J. Chem. Phys.}\ }\textbf {\bibinfo {volume} {156}},\ \bibinfo
  {pages} {174109} (\bibinfo {year} {2022})}\BibitemShut {NoStop}%
\bibitem [{\citenamefont {Han}\ \emph {et~al.}(2023)\citenamefont {Han},
  \citenamefont {Ha},\ and\ \citenamefont {Min}}]{HanJCTC2023}%
  \BibitemOpen
  \bibfield  {author} {\bibinfo {author} {\bibfnamefont {D.}~\bibnamefont
  {Han}}, \bibinfo {author} {\bibfnamefont {J.-K.}\ \bibnamefont {Ha}},\ and\
  \bibinfo {author} {\bibfnamefont {S.~K.}\ \bibnamefont {Min}},\ }\bibfield
  {title} {\bibinfo {title} {{Real-Space and Real-Time Propagation for
  Correlated Electron–Nuclear Dynamics Based on Exact Factorization}},\
  }\href {https://doi.org/10.1021/acs.jctc.2c00939} {\bibfield  {journal}
  {\bibinfo  {journal} {J. Chem. Theory Comput.}\ }\textbf {\bibinfo {volume}
  {19}},\ \bibinfo {pages} {2186} (\bibinfo {year} {2023})}\BibitemShut
  {NoStop}%
\bibitem [{\citenamefont {Dupuy}\ \emph {et~al.}(2024)\citenamefont {Dupuy},
  \citenamefont {Rikus},\ and\ \citenamefont {Maitra}}]{DupuyJPCL2024}%
  \BibitemOpen
  \bibfield  {author} {\bibinfo {author} {\bibfnamefont {L.}~\bibnamefont
  {Dupuy}}, \bibinfo {author} {\bibfnamefont {A.}~\bibnamefont {Rikus}},\ and\
  \bibinfo {author} {\bibfnamefont {N.~T.}\ \bibnamefont {Maitra}},\ }\bibfield
   {title} {\bibinfo {title} {{Exact-Factorization-Based Surface Hopping
  without Velocity Adjustment}},\ }\href
  {https://doi.org/10.1021/acs.jpclett.4c00115} {\bibfield  {journal} {\bibinfo
   {journal} {J. Phys. Chem. Lett.}\ }\textbf {\bibinfo {volume} {15}},\
  \bibinfo {pages} {2643} (\bibinfo {year} {2024})}\BibitemShut {NoStop}%
\bibitem [{\citenamefont {Han}\ \emph {et~al.}(2025{\natexlab{a}})\citenamefont
  {Han}, \citenamefont {Martens},\ and\ \citenamefont {Akimov}}]{HanJCTC2025}%
  \BibitemOpen
  \bibfield  {author} {\bibinfo {author} {\bibfnamefont {D.}~\bibnamefont
  {Han}}, \bibinfo {author} {\bibfnamefont {C.~C.}\ \bibnamefont {Martens}},\
  and\ \bibinfo {author} {\bibfnamefont {A.~V.}\ \bibnamefont {Akimov}},\
  }\bibfield  {title} {\bibinfo {title} {{Generalization of Quantum-Trajectory
  Surface Hopping to Multiple Quantum States}},\ }\href
  {https://doi.org/10.1021/acs.jctc.4c01751} {\bibfield  {journal} {\bibinfo
  {journal} {J. Chem. Theory Comput.}\ }\textbf {\bibinfo {volume} {21}},\
  \bibinfo {pages} {2839} (\bibinfo {year} {2025}{\natexlab{a}})}\BibitemShut
  {NoStop}%
\bibitem [{\citenamefont {Han}\ \emph {et~al.}(2025{\natexlab{b}})\citenamefont
  {Han}, \citenamefont {Lee},\ and\ \citenamefont {Min}}]{HanJCTC2025_2}%
  \BibitemOpen
  \bibfield  {author} {\bibinfo {author} {\bibfnamefont {D.}~\bibnamefont
  {Han}}, \bibinfo {author} {\bibfnamefont {J.~H.}\ \bibnamefont {Lee}},\ and\
  \bibinfo {author} {\bibfnamefont {S.~K.}\ \bibnamefont {Min}},\ }\bibfield
  {title} {\bibinfo {title} {Orbital-based correlated electron–nuclear
  dynamics for extended systems with exact factorization},\ }\href
  {https://doi.org/10.1021/acs.jctc.5c01575} {\bibfield  {journal} {\bibinfo
  {journal} {J. Chem. Theory Comput.}\ }\textbf {\bibinfo {volume} {21}},\
  \bibinfo {pages} {11415} (\bibinfo {year} {2025}{\natexlab{b}})}\BibitemShut
  {NoStop}%
\bibitem [{\citenamefont {Arribas}\ and\ \citenamefont
  {Maitra}(2024)}]{ArribasPRL2024}%
  \BibitemOpen
  \bibfield  {author} {\bibinfo {author} {\bibfnamefont {E.~V.}\ \bibnamefont
  {Arribas}}\ and\ \bibinfo {author} {\bibfnamefont {N.~T.}\ \bibnamefont
  {Maitra}},\ }\bibfield  {title} {\bibinfo {title} {{Electronic Coherences in
  Molecules: The Projected Nuclear Quantum Momentum as a Hidden Agent}},\
  }\href {https://doi.org/10.1103/PhysRevLett.133.233201} {\bibfield  {journal}
  {\bibinfo  {journal} {Phys. Rev. Lett.}\ }\textbf {\bibinfo {volume} {133}},\
  \bibinfo {pages} {233201} (\bibinfo {year} {2024})}\BibitemShut {NoStop}%
\end{thebibliography}%


\begin{thebibliography}{12}%
\makeatletter
\providecommand \@ifxundefined [1]{%
 \@ifx{#1\undefined}
}%
\providecommand \@ifnum [1]{%
 \ifnum #1\expandafter \@firstoftwo
 \else \expandafter \@secondoftwo
 \fi
}%
\providecommand \@ifx [1]{%
 \ifx #1\expandafter \@firstoftwo
 \else \expandafter \@secondoftwo
 \fi
}%
\providecommand \natexlab [1]{#1}%
\providecommand \enquote  [1]{``#1''}%
\providecommand \bibnamefont  [1]{#1}%
\providecommand \bibfnamefont [1]{#1}%
\providecommand \citenamefont [1]{#1}%
\providecommand \href@noop [0]{\@secondoftwo}%
\providecommand \href [0]{\begingroup \@sanitize@url \@href}%
\providecommand \@href[1]{\@@startlink{#1}\@@href}%
\providecommand \@@href[1]{\endgroup#1\@@endlink}%
\providecommand \@sanitize@url [0]{\catcode `\\12\catcode `\$12\catcode
  `\&12\catcode `\#12\catcode `\^12\catcode `\_12\catcode `\%12\relax}%
\providecommand \@@startlink[1]{}%
\providecommand \@@endlink[0]{}%
\providecommand \url  [0]{\begingroup\@sanitize@url \@url }%
\providecommand \@url [1]{\endgroup\@href {#1}{\urlprefix }}%
\providecommand \urlprefix  [0]{URL }%
\providecommand \Eprint [0]{\href }%
\providecommand \doibase [0]{https://doi.org/}%
\providecommand \selectlanguage [0]{\@gobble}%
\providecommand \bibinfo  [0]{\@secondoftwo}%
\providecommand \bibfield  [0]{\@secondoftwo}%
\providecommand \translation [1]{[#1]}%
\providecommand \BibitemOpen [0]{}%
\providecommand \bibitemStop [0]{}%
\providecommand \bibitemNoStop [0]{.\EOS\space}%
\providecommand \EOS [0]{\spacefactor3000\relax}%
\providecommand \BibitemShut  [1]{\csname bibitem#1\endcsname}%
\let\auto@bib@innerbib\@empty
\bibitem [{\citenamefont {Arribas}\ and\ \citenamefont
  {Maitra}(2024)}]{ArribasPRL2024}%
  \BibitemOpen
  \bibfield  {author} {\bibinfo {author} {\bibfnamefont {E.~V.}\ \bibnamefont
  {Arribas}}\ and\ \bibinfo {author} {\bibfnamefont {N.~T.}\ \bibnamefont
  {Maitra}},\ }\bibfield  {title} {\bibinfo {title} {{Electronic Coherences in
  Molecules: The Projected Nuclear Quantum Momentum as a Hidden Agent}},\
  }\href {https://doi.org/10.1103/PhysRevLett.133.233201} {\bibfield  {journal}
  {\bibinfo  {journal} {Phys. Rev. Lett.}\ }\textbf {\bibinfo {volume} {133}},\
  \bibinfo {pages} {233201} (\bibinfo {year} {2024})}\BibitemShut {NoStop}%
\bibitem [{\citenamefont {Agostini}\ \emph {et~al.}(2016)\citenamefont
  {Agostini}, \citenamefont {Min}, \citenamefont {Abedi},\ and\ \citenamefont
  {Gross}}]{AgostiniJCTC2016}%
  \BibitemOpen
  \bibfield  {author} {\bibinfo {author} {\bibfnamefont {F.}~\bibnamefont
  {Agostini}}, \bibinfo {author} {\bibfnamefont {S.~K.}\ \bibnamefont {Min}},
  \bibinfo {author} {\bibfnamefont {A.}~\bibnamefont {Abedi}},\ and\ \bibinfo
  {author} {\bibfnamefont {E.~K.~U.}\ \bibnamefont {Gross}},\ }\bibfield
  {title} {\bibinfo {title} {{Quantum-Classical Nonadiabatic Dynamics: Coupled-
  vs Independent-Trajectory Methods}},\ }\href
  {https://doi.org/10.1021/acs.jctc.5b01180} {\bibfield  {journal} {\bibinfo
  {journal} {J. Chem. Theory Comput.}\ }\textbf {\bibinfo {volume} {12}},\
  \bibinfo {pages} {2127} (\bibinfo {year} {2016})}\BibitemShut {NoStop}%
\bibitem [{\citenamefont {Min}\ \emph {et~al.}(2015)\citenamefont {Min},
  \citenamefont {Agostini},\ and\ \citenamefont {Gross}}]{MinPRL2015}%
  \BibitemOpen
  \bibfield  {author} {\bibinfo {author} {\bibfnamefont {S.~K.}\ \bibnamefont
  {Min}}, \bibinfo {author} {\bibfnamefont {F.}~\bibnamefont {Agostini}},\ and\
  \bibinfo {author} {\bibfnamefont {E.~K.~U.}\ \bibnamefont {Gross}},\
  }\bibfield  {title} {\bibinfo {title} {{Coupled-Trajectory Quantum-Classical
  Approach to Electronic Decoherence in Nonadiabatic Processes}},\ }\href
  {https://doi.org/10.1103/PhysRevLett.115.073001} {\bibfield  {journal}
  {\bibinfo  {journal} {Phys. Rev. Lett.}\ }\textbf {\bibinfo {volume} {115}},\
  \bibinfo {pages} {073001} (\bibinfo {year} {2015})}\BibitemShut {NoStop}%
\bibitem [{\citenamefont {Ha}\ and\ \citenamefont {Min}(2022)}]{HaJCP2022}%
  \BibitemOpen
  \bibfield  {author} {\bibinfo {author} {\bibfnamefont {J.-K.}\ \bibnamefont
  {Ha}}\ and\ \bibinfo {author} {\bibfnamefont {S.~K.}\ \bibnamefont {Min}},\
  }\bibfield  {title} {\bibinfo {title} {Independent trajectory mixed
  quantum-classical approaches based on the exact factorization},\ }\href
  {https://doi.org/10.1063/5.0084493} {\bibfield  {journal} {\bibinfo
  {journal} {J. Chem. Phys.}\ }\textbf {\bibinfo {volume} {156}},\ \bibinfo
  {pages} {174109} (\bibinfo {year} {2022})}\BibitemShut {NoStop}%
\bibitem [{\citenamefont {Arribas}\ and\ \citenamefont
  {Maitra}(2023)}]{ArribasJCP2023}%
  \BibitemOpen
  \bibfield  {author} {\bibinfo {author} {\bibfnamefont {E.~V.}\ \bibnamefont
  {Arribas}}\ and\ \bibinfo {author} {\bibfnamefont {N.~T.}\ \bibnamefont
  {Maitra}},\ }\bibfield  {title} {\bibinfo {title} {Energy-conserving coupled
  trajectory mixed quantum–classical dynamics},\ }\href
  {https://doi.org/10.1063/5.0149116} {\bibfield  {journal} {\bibinfo
  {journal} {J. Chem. Phys.}\ }\textbf {\bibinfo {volume} {158}},\ \bibinfo
  {pages} {161105} (\bibinfo {year} {2023})}\BibitemShut {NoStop}%
\bibitem [{\citenamefont {Tully}(1990)}]{TullyJCP1990}%
  \BibitemOpen
  \bibfield  {author} {\bibinfo {author} {\bibfnamefont {J.~C.}\ \bibnamefont
  {Tully}},\ }\bibfield  {title} {\bibinfo {title} {Molecular dynamics with
  electronic transitions},\ }\href {https://doi.org/10.1063/1.459170}
  {\bibfield  {journal} {\bibinfo  {journal} {J. Chem. Phys.}\ }\textbf
  {\bibinfo {volume} {93}},\ \bibinfo {pages} {1061} (\bibinfo {year}
  {1990})}\BibitemShut {NoStop}%
\bibitem [{\citenamefont {Ha}\ \emph {et~al.}(2018)\citenamefont {Ha},
  \citenamefont {Lee},\ and\ \citenamefont {Min}}]{HaJPCL2018}%
  \BibitemOpen
  \bibfield  {author} {\bibinfo {author} {\bibfnamefont {J.-K.}\ \bibnamefont
  {Ha}}, \bibinfo {author} {\bibfnamefont {I.~S.}\ \bibnamefont {Lee}},\ and\
  \bibinfo {author} {\bibfnamefont {S.~K.}\ \bibnamefont {Min}},\ }\bibfield
  {title} {\bibinfo {title} {{Surface Hopping Dynamics beyond Nonadiabatic
  Couplings for Quantum Coherence}},\ }\href
  {https://doi.org/10.1021/acs.jpclett.8b00060} {\bibfield  {journal} {\bibinfo
   {journal} {J. Phys. Chem. Lett.}\ }\textbf {\bibinfo {volume} {9}},\
  \bibinfo {pages} {1097} (\bibinfo {year} {2018})}\BibitemShut {NoStop}%
\bibitem [{\citenamefont {Shenvi}\ \emph {et~al.}(2011)\citenamefont {Shenvi},
  \citenamefont {Subotnik},\ and\ \citenamefont {Yang}}]{ShenviJCP2011}%
  \BibitemOpen
  \bibfield  {author} {\bibinfo {author} {\bibfnamefont {N.}~\bibnamefont
  {Shenvi}}, \bibinfo {author} {\bibfnamefont {J.~E.}\ \bibnamefont
  {Subotnik}},\ and\ \bibinfo {author} {\bibfnamefont {W.}~\bibnamefont
  {Yang}},\ }\bibfield  {title} {\bibinfo {title} {{Phase-corrected surface
  hopping: Correcting the phase evolution of the electronic wavefunction}},\
  }\href {https://doi.org/10.1063/1.3603447} {\bibfield  {journal} {\bibinfo
  {journal} {J. Chem. Phys.}\ }\textbf {\bibinfo {volume} {135}},\ \bibinfo
  {pages} {024101} (\bibinfo {year} {2011})}\BibitemShut {NoStop}%
\bibitem [{\citenamefont {Feit}\ \emph {et~al.}(1982)\citenamefont {Feit},
  \citenamefont {Fleck},\ and\ \citenamefont {Steiger}}]{FeitJCP1982}%
  \BibitemOpen
  \bibfield  {author} {\bibinfo {author} {\bibfnamefont {M.}~\bibnamefont
  {Feit}}, \bibinfo {author} {\bibfnamefont {J.}~\bibnamefont {Fleck}},\ and\
  \bibinfo {author} {\bibfnamefont {A.}~\bibnamefont {Steiger}},\ }\bibfield
  {title} {\bibinfo {title} {Solution of the {S}chrödinger equation by a
  spectral method},\ }\href {https://doi.org/10.1016/0021-9991(82)90091-2}
  {\bibfield  {journal} {\bibinfo  {journal} {J. Comput. Phys.}\ }\textbf
  {\bibinfo {volume} {47}},\ \bibinfo {pages} {412} (\bibinfo {year}
  {1982})}\BibitemShut {NoStop}%
\bibitem [{\citenamefont {Granucci}\ and\ \citenamefont
  {Persico}(2007)}]{GranucciJCP2007}%
  \BibitemOpen
  \bibfield  {author} {\bibinfo {author} {\bibfnamefont {G.}~\bibnamefont
  {Granucci}}\ and\ \bibinfo {author} {\bibfnamefont {M.}~\bibnamefont
  {Persico}},\ }\bibfield  {title} {\bibinfo {title} {Critical appraisal of the
  fewest switches algorithm for surface hopping},\ }\href
  {https://doi.org/10.1063/1.2715585} {\bibfield  {journal} {\bibinfo
  {journal} {J. Chem. Phys.}\ }\textbf {\bibinfo {volume} {126}},\ \bibinfo
  {pages} {134114} (\bibinfo {year} {2007})}\BibitemShut {NoStop}%
\bibitem [{\citenamefont {Xu}\ and\ \citenamefont {Wang}(2019)}]{XuJCP2019}%
  \BibitemOpen
  \bibfield  {author} {\bibinfo {author} {\bibfnamefont {J.}~\bibnamefont
  {Xu}}\ and\ \bibinfo {author} {\bibfnamefont {L.}~\bibnamefont {Wang}},\
  }\bibfield  {title} {\bibinfo {title} {{Branching corrected surface hopping:
  Resetting wavefunction coefficients based on judgement of wave packet
  reflection}},\ }\href {https://doi.org/10.1063/1.5090927} {\bibfield
  {journal} {\bibinfo  {journal} {J. Chem. Phys.}\ }\textbf {\bibinfo {volume}
  {150}},\ \bibinfo {pages} {164101} (\bibinfo {year} {2019})}\BibitemShut
  {NoStop}%
\bibitem [{\citenamefont {Shao}\ \emph {et~al.}(2023)\citenamefont {Shao},
  \citenamefont {Shi}, \citenamefont {Xu},\ and\ \citenamefont
  {Wang}}]{ShaoJPCL2023}%
  \BibitemOpen
  \bibfield  {author} {\bibinfo {author} {\bibfnamefont {C.}~\bibnamefont
  {Shao}}, \bibinfo {author} {\bibfnamefont {Z.}~\bibnamefont {Shi}}, \bibinfo
  {author} {\bibfnamefont {J.}~\bibnamefont {Xu}},\ and\ \bibinfo {author}
  {\bibfnamefont {L.}~\bibnamefont {Wang}},\ }\bibfield  {title} {\bibinfo
  {title} {Learning decoherence time formulas for surface hopping from quantum
  dynamics},\ }\href {https://doi.org/10.1021/acs.jpclett.3c02019} {\bibfield
  {journal} {\bibinfo  {journal} {J. Phys. Chem. Lett.}\ }\textbf {\bibinfo
  {volume} {14}},\ \bibinfo {pages} {7680} (\bibinfo {year}
  {2023})}\BibitemShut {NoStop}%
\end{thebibliography}%
\end{document}


\title{Supporting Information for "Unifying Decoherence and Phase Evolution in Mixed Quantum–Classical Dynamics through Exact Factorization"
}

\date{\today}
\author{Jong-Kwon Ha}
\affiliation{Department of Chemistry, Dalhousie University, 6274 Coburg Rd, Halifax, NS B3H 4R2, Canada}
\author{Seong Ho Kim}
\affiliation{Department of Chemistry, Ulsan National Institute of Science and Technology (UNIST), 50 UNIST-gil, Ulju-gun, Ulsan 44919, South Korea}
\author{Seung Kyu Min}
\email{skmin@unist.ac.kr}
\affiliation{Department of Chemistry, Ulsan National Institute of Science and Technology (UNIST), 50 UNIST-gil, Ulju-gun, Ulsan 44919, South Korea}

\maketitle
\section{Exact mixed Quantum-classical equations of motion in the Born-Oppenheimer representation}
\subsection{Electronic time-dependent Schr\"{o}dinger equation}
If we expand the electronic state with the Born-Oppenheimer (BO) basis states in the electronic time-dependent Schr\"{o}dinger equation (TDSE) for the exact mixed quantum-classical (MQC) dynamics, we get 
\begin{subequations}
\begin{align}
    \dfrac{d}{dt}&C^{(I)}_j=-\dfrac{i}{\hbar}\epsilon^{(I)}_jC^{(I)}_j- \sum_k\sum_\nu\dfrac{\bP^{(I)}_\nu}{M_\nu}\cdot\bd^{(I)}_{\nu,jk}C^{(I)}_k-\dfrac{i}{\hbar}\tilde{\epsilon}^{(I)}C^{(I)}_j\label{eq:eh_1}\\
    &+\sum_\nu\dfrac{i\hbar}{2M_\nu}\bm{\mathcal{P}}^{(I)}_\nu\cdot\left(\nabla_\nu C_j+\sum_{k}C_k\bd_{\nu,jk}-\dfrac{i}{\hbar}\bA_\nu C_j\right)^{(I)}\label{eq:enc1_1}\\
    &+\sum_\nu\dfrac{i\hbar}{2M_\nu}\left(\nabla^2_\nu C_j+2\sum_k\nabla_\nu C_k\bd_{\nu,jk}+\sum_k C_kg_{\nu,jk}-2\dfrac{i}{\hbar}\bA_\nu\cdot\nabla_\nu C_j-2\dfrac{i}{\hbar}\bA_\nu\cdot\sum_kC_k\bd_{\nu,jk}\right)^{(I)}\label{eq:enc2-1_1}\\
    &+\sum_\nu\dfrac{i\hbar}{2M_\nu}\left(\sum_k\nabla_\nu C^*_kC_k+2\sum_{kl} \nabla_\nu C^*_kC_l\bd^*_{\nu,lk}+\sum_{kl}\nabla_\nu C^*_kC_lg^*_{\nu,lk}\right)^{(I)} C^{(I)}_j\label{eq:enc2-1_2}\\
    &+\sum_\nu\dfrac{i\hbar}{2M_\nu}\left(\sum_k|\nabla_\nu C_k|^2+\sum_{kl}\nabla_\nu C^*_kC_l\bd_{\nu,kl}+\sum_{kl} C^*_k\nabla_\nu C_l\bd^*_{\nu,lk}+\sum_{kl}C^*_kC_l\braket{\nabla_\nu\phi_k}{\nabla_\nu\phi_l}-\dfrac{1}{\hbar^2}|\bA_\nu|^2\right)^{(I)}C^{(I)}_j\label{eq:enc2-2},
\end{align}
\end{subequations}
where $\bd_{\nu,kl}=\braket{\phi_k}{\nabla_\nu\phi_l}_e$ and $g_{\nu,kl}=\braket{\phi_k}{\nabla^2_\nu\phi_l}_e$ are first and second order nonadiabatic couplings (NACs).
The line (\ref{eq:eh_1}) contains the Ehrenfest term plus time-dependent scalar potential contribution, (\ref{eq:enc1_1}) is from $\hH^{(1)}_\mathrm{ENC}$, (\ref{eq:enc2-1_1}) and (\ref{eq:enc2-1_2}) are from $\nabla^2_\nu\hat{\Gamma}_\dulR$ term of $\hH^{(2)}_\mathrm{ENC}$, and (\ref{eq:enc2-2}) is from $\nabla_\nu\hat{\Gamma}_\dulR\cdot\nabla_\nu\hat{\Gamma}_\dulR$ term of $\hH^{(2)}_\mathrm{ENC}$.
Neglecting NACs and writing the BO coefficients in polar form, we can write the equation as
\begin{equation}
\begin{split}
    \dfrac{d}{dt}&C^{(I)}_j=-\dfrac{i}{\hbar}\epsilon^{(I)}_jC^{(I)}_j- \sum_k\sum_\nu\dfrac{\bP^{(I)}_\nu}{M_\nu}\cdot\bd^{(I)}_{\nu,jk}C^{(I)}_k-\dfrac{i}{\hbar}\tilde{\epsilon}^{(I)}C^{(I)}_j\\
    &-\sum_\nu\dfrac{1}{2M_\nu}\bm{\mathcal{P}}^{(I)}_\nu\cdot\left(\nabla_\nu S_j - \bA_\nu\right)^{(I)}C^{(I)}_j
    +\sum_\nu\dfrac{i\hbar}{2M_\nu}\bm{\mathcal{P}}^{(I)}_\nu\cdot\dfrac{\nabla_\nu|C_j|}{|C_j|}C_j\\
    &-\sum_\nu\dfrac{1}{2M_\nu}\sum_k\left(2\dfrac{\nabla_\nu|C_j|}{|C_j|}\cdot\nabla_\nu S_j+\nabla^2_\nu S_j-2\dfrac{\nabla_\nu|C_j|}{|C_j|}\cdot\bA_\nu-\dfrac{\nabla_\nu|C_k|}{|C_k|}\cdot\nabla_\nu S_k-\nabla^2_\nu S_k\right)^{(I)}|C^{(I)}_k|^2C^{(I)}_j\\
    &+\sum_\nu\dfrac{i\hbar}{2M_\nu}\sum_k\left(\dfrac{\nabla^2_\nu|C_j|}{|C_j|}-\dfrac{1}{\hbar^2}|\nabla_\nu S_j|^2+\dfrac{2}{\hbar^2}\nabla_\nu S_j\cdot\bA_\nu+\sum_k\dfrac{\nabla^2_\nu|C_k|}{|C_k|}-\dfrac{1}{\hbar^2}|\nabla_\nu S_k|^2\right)^{(I)}|C^{(I)}_k|^2C^{(I)}_j\\
    &+\sum_\nu\dfrac{i\hbar}{2M_\nu}\sum_k\left(\dfrac{\nabla_\nu|C_k|}{|C_k|}\cdot\dfrac{\nabla_\nu|C_k|}{|C_k|}+\dfrac{1}{\hbar^2}|\nabla_\nu S_k|^2-\dfrac{1}{\hbar^2}|\bA_\nu|^2\right)^{(I)}|C^{(I)}_k|^2C^{(I)}_j.
\end{split}
\end{equation}
Using $\sum_k|C_k|=1$, $\bA_\nu = \sum_k|C_k|^2\nabla_\nu S_k+\hbar\Im\{C^*_kC_l\bd_{\nu,kl}\}$, and neglecting NAC for $\bA_\nu$, we get
\begin{equation}
\begin{split}
    \dfrac{d}{dt}&C^{(I)}_j=-\dfrac{i}{\hbar}\epsilon^{(I)}_jC^{(I)}_j- \sum_k\sum_\nu\dfrac{\bP^{(I)}_\nu}{M_\nu}\cdot\bd^{(I)}_{\nu,jk}C^{(I)}_k-\dfrac{i}{\hbar}\tilde{\epsilon}^{(I)}C^{(I)}_j\\
    &-\sum_\nu\dfrac{1}{2M_\nu}\sum_k\left(\bm{\mathcal{P}}^{(I)}_\nu+\bm{\mathcal{Q}}^{(I)}_{\nu,jk}+\nabla_\nu\right)\cdot\mathbf{D}_{\nu,jk}^{(I)}|C^{(I)}_k|^2C^{(I)}_j\\
    &+\sum_\nu\dfrac{i\hbar}{2M_\nu}\sum_k\left(\dfrac{\nabla^2_\nu|C_j|}{|C_j|}+\bm{\mathcal{P}}^{(I)}_\nu\cdot\dfrac{\nabla_\nu|C_j|}{|C_j|}-\dfrac{1}{\hbar^2}|\nabla_\nu S_j-\bA_\nu|^2\right)^{(I)}C^{(I)}_j,
\end{split}
\end{equation}
where $\mathbf{D}_{\nu,jk} = \nabla_\nu S_j - \nabla_\nu S_k$, $\bm{\mathcal{P}}_\nu=\nabla_\nu|\chi|^2/|\chi|^2$ is the total nuclear quantum momentum (QM), and $\bm{\mathcal{Q}}_{\nu,jk}=\nabla_\nu|C_j|^2/|C_j|^2+\nabla_\nu|C_k|^2/|C_k|^2$ is the reduced projected quantum momentum (PQM)~\cite{ArribasPRL2024}.
As a result, the electron-nuclear correlation (ENC) Hamiltonian matrix is diagonal in the absence of NAC, and its real and imaginary part corresponds to decoherence and phase correction, respectively.
We note that the new decoherence correction term originates solely from $\nabla^2_\nu\hat{\Gamma}_\dulR$ where $\hat{\Gamma}_\dulR=\ket{\Phi_\dulR}\bra{\Phi_\dulR}$ is the reduced electron density operator, while $\nabla_\nu\hat{\Gamma}_\dulR\cdot\nabla_\nu\hat{\Gamma}_\dulR$ contributes merely to the global phase with cancellations with the phase correction term from $\nabla^2_\nu\hat{\Gamma}_\dulR$ that just make the equation simpler.

Finally, keeping the lowest order of $\hbar$ terms, we get the final equation:
\begin{equation}\label{eq:tdse_bo}
\begin{split}
    \dfrac{d}{dt}C^{(I)}_j=&-\dfrac{i}{\hbar}\epsilon^{(I)}_jC^{(I)}_j- \sum_k\sum_\nu\dfrac{\bP^{(I)}_\nu}{M_\nu}\cdot\bd^{(I)}_{\nu,jk}C^{(I)}_k\\
    &-\sum_\nu\dfrac{1}{2M_\nu}\sum_k\left(\bm{\mathcal{P}}^{(I)}_\nu+\bm{\mathcal{Q}}^{(I)}_{\nu,jk}+\nabla_\nu\right)\cdot\mathbf{D}_{\nu,jk}^{(I)}|C^{(I)}_k|^2C^{(I)}_j\\
    &-\dfrac{i}{\hbar}\left(\sum_\nu\dfrac{|\nabla_\nu S^{(I)}_j-\bA^{(I)}_\nu|^2}{2M_\nu}+\tilde{\epsilon}^{(I)}\right)C^{(I)}_j\\
    \equiv&-\dfrac{i}{\hbar}\epsilon^{(I)}_jC^{(I)}_j- \sum_k\sum_\nu\dfrac{\bP^{(I)}_\nu}{M_\nu}\cdot\bd^{(I)}_{\nu,jk}C^{(I)}_k\\
    &+\left(\dfrac{1}{\hbar}H^\mathrm{DC}_j-\dfrac{i}{\hbar}H^\mathrm{PC}_j+\right)C^{(I)}_j
\end{split}
\end{equation}
for the electronic TDSE in the BO representation.

\subsection{Nuclear force}
The nuclear force is
\begin{equation}
\begin{split}
    \bF^{(I)}_\nu(t) =& -\nabla_\nu \tilde{\epsilon}^{(I)} +\dot{\bA}^{(I)}_\nu\\
            =&-\nabla_\nu\left(\bra{\ewf}\hat{H}_\mathrm{BO}\ket{\ewf}_e+\sum_\mu\dfrac{\hbar^2\braket{\nabla_\mu\ewf}{\nabla_\mu\ewf}_e-|\bA_\mu|^2}{2M_\mu}\right)^{(I)}
            + 2\Re\left(i\hbar\braket{\nabla_\nu\ewf}{\frac{d}{dt}\ewf}_e\right)^{(I)}\\
            =&-\nabla_\nu\left(\bra{\ewf}\hat{H}_\mathrm{BO}\ket{\ewf}_e+\sum_\mu\dfrac{\hbar^2\braket{\nabla_\mu\ewf}{\nabla_\mu\ewf}_e-|\bA_\mu|^2}{2M_\mu}\right)^{(I)} +2\Re\left(\bra{\nabla_\nu\ewf}(\hat{H}_\mathrm{BO}+\hat{H}_\mathrm{ENC}-\tilde{\epsilon})\ket{\ewf}_e\right)\\
            =&-\bra{\ewf}\nabla_\nu\hat{H}_\mathrm{BO}\ket{\ewf}^{(I)}_e-\nabla_\nu\left(\sum_\mu\dfrac{\hbar^2\braket{\nabla_\mu\ewf}{\nabla_\mu\ewf}_e-|\bA_\mu|^2}{2M_\mu}\right)^{(I)}+2\Re\left(\bra{\nabla_\nu\ewf}\hat{H}_\mathrm{ENC}\ket{\ewf}_e\right)^{(I)}.
\end{split}
\end{equation}
The full nuclear force, including all NAC contributions within the ENC formalism, can be derived by straightforward but lengthy algebra involving numerous terms.
For brevity, and because the NACs are generally highly localized in the nuclear configuration space, these contributions are neglected in the following.
As a result, the $H_\mathrm{ENC}$ is diagonal in the BO expansion in the absence of NACs, the last term in the above equation becomes
\begin{equation}
\begin{split}
    2\Re\Bigg(\sum_{k}\nabla_\nu C^{*}_kH^\mathrm{ENC}_kC_k\Bigg)^{(I)}=&2\Re\left\{\sum_{k}\nabla_\nu C^{*}_k\left(i H^\mathrm{DC}_k+H^\mathrm{PC}_k\right)C_k\right\}^{(I)}\\
    =&2\Re\Bigg\{\sum_k\left(\dfrac{\nabla_\nu|C_k|}{|C_k|}-\dfrac{i}{\hbar}\nabla_\nu S_k\right)\left(i H^\mathrm{DC}_k+H^\mathrm{PC}_k\right)|C_k|^2\Bigg\}^{(I)}\\
    =&2\left\{\sum_k\left(\dfrac{1}{\hbar}\nabla_\nu S_kH^\mathrm{DC}_k+\dfrac{\nabla_\nu|C_k|}{|C_k|}H^\mathrm{PC}_k\right)|C_k|^2\right\}^{(I)}.
\end{split}
\end{equation}

Then, the nuclear force in the BO representation becomes
\begin{equation}
\begin{split}
    \bF_\nu=&-\sum_k|C_k|^2\nabla_\nu\epsilon_k - \sum_{k,l}C_kC_l^*(\epsilon_k-\epsilon_l)\bd_{\nu,lk}-\nabla_\nu\left(\sum_\mu\dfrac{\hbar^2\braket{\nabla_\mu\ewf}{\nabla_\mu\ewf}_e-|\bA_\mu|^2}{2M_\mu}\right)\\
        &-\sum_{k,l}\mathbf{D}_{\nu,kl}\left\{\sum_\mu\dfrac{1}{2M_\mu}\left(
    \bm{\mathcal{P}}_{\mu}+\bm{\mathcal{Q}}_{\mu,kl}+\nabla_\mu\right)
    \cdot\mathbf{D}_{\mu,kl}\right\}|C_l|^2|C_k|^2\\
    &+2\sum_{k}\dfrac{\nabla_\nu|C_k|}{|C_k|}\left(-\tilde{\epsilon}+\sum_\mu\dfrac{|\nabla_\mu S_k-\bA_\mu|^2}{2M_\mu}\right)|C_k|^2.
\end{split}
\end{equation}
We omit $(I)$ for brevity in this subsection.
Since $\sum_k|C_k|\nabla_\nu|C_k|=\nabla_\nu(\sum_k|C_k|^2)/2 = 0$, $\sum_k|C_k|\nabla_\nu|C_k|\tilde{\epsilon}=0$ meaning that there is no contribution from the global phase on the nuclear force,
\begin{equation}
\begin{split}
    \bF_\nu=&-\sum_k|C_k|^2\nabla_\nu\epsilon_k - \sum_{k,l}C_kC_l^*(\epsilon_k-\epsilon_l)\bd_{\nu,lk}-\nabla_\nu\left(\sum_\mu\dfrac{\hbar^2\braket{\nabla_\mu\ewf}{\nabla_\mu\ewf}_e-|\bA_\mu|^2}{2M_\mu}\right)\\
        &-\sum_{k,l}\mathbf{D}_{\nu,kl}\left\{\sum_\mu\dfrac{1}{2M_\mu}\left(
    \bm{\mathcal{P}}_{\mu}+\bm{\mathcal{Q}}_{\mu,kl}+\nabla_\mu\right)
    \cdot\mathbf{D}_{\mu,kl}\right\}|C_l|^2|C_k|^2\\
    &+\sum_{k}\dfrac{\nabla_\nu|C_k|}{|C_k|}\left(\sum_\mu\dfrac{|\nabla_\mu S_k-\bA_\mu|^2}{M_\mu}\right)|C_k|^2
\end{split}
\end{equation}
The last term from PC can be rearranged
\begin{equation}
\begin{split}
    \sum_k\dfrac{\nabla_\nu|C_k|}{|C_k|}\left(\sum_\mu\dfrac{|\nabla_\mu S_k-\bA_\mu|^2}{M_\mu}\right)&|C_k|^2
    =\sum_k\dfrac{\nabla_\nu|C_k|}{|C_k|}\left(\sum_\mu\dfrac{|\nabla_\mu S_k|^2+|\bA_\mu|^2-2\nabla_\mu S_k\cdot\bA_\mu}{M_\mu}\right)|C_k|^2\\
    =&\sum_k\dfrac{\nabla_\nu|C_k|}{|C_k|}\left(\sum_\mu\dfrac{|\nabla_\mu S_k|^2-2\nabla_\mu S_k\cdot\bA_\mu}{M_\mu}\right)|C_k|^2\\
    =&\sum_k\dfrac{\nabla_\nu|C_k|}{|C_k|}\left(\sum_\mu\dfrac{|\nabla_\mu S_k|^2+\sum_l|C_l|^2|\nabla_\mu S_l|^2-2\nabla_\mu S_k\cdot\sum_l|C_l|^2\nabla_\mu S_l}{M_\mu}\right)|C_k|^2\\
    =&\sum_{k,l}\dfrac{\nabla_\nu|C_k|}{|C_k|}\left(\sum_\mu\dfrac{|\nabla_\mu S_k-\nabla_\mu S_l|^2}{M_\mu}\right)|C_k|^2|C_l|^2,
\end{split}
\end{equation}
where we use $\sum_k|C_k|^2=1$ and $\sum_k|C_k|\nabla_\nu|C_k|=\nabla_\nu(\sum_k|C_k|^2)/2 = 0$, and neglected NAC in $\bA_\mu$ again.
As a result,
\begin{equation}
\begin{split}
    \bF_\nu=&-\sum_k|C_k|^2\nabla_\nu\epsilon_k - \sum_{k,l}C_kC_l^*(\epsilon_k-\epsilon_l)\bd_{\nu,lk}-\nabla_\nu\left(\sum_\mu\dfrac{\hbar^2\braket{\nabla_\mu\ewf}{\nabla_\mu\ewf}_e-|\bA_\mu|^2}{2M_\mu}\right)\\
    &-\sum_{k,l}\mathbf{D}_{\nu,kl}\left\{\sum_\mu\dfrac{1}{2M_\mu}\left(
    \bm{\mathcal{P}}_{\mu}+\bm{\mathcal{Q}}_{\mu,kl}+\nabla_\mu\right)
    \cdot\mathbf{D}_{\mu,kl}\right\}|C_l|^2|C_k|^2\\
    &+\sum_{k,l}\dfrac{\nabla_\nu|C_k|}{|C_k|}\left(\sum_\mu\dfrac{|\mathbf{D}_{\mu,kl}|^2}{M_\mu}\right)|C_k|^2|C_l|^2,
\end{split}
\end{equation}
where the Ehrenfest force is given by $\mathbf{F}_\nu^\mathrm{Eh} = -\sum_k|C_k|^2\nabla_\nu\epsilon_k -\sum_{k,l}C_kC_l^*(\epsilon_k-\epsilon_l)\bd_{\nu,lk}$, the force from the QM and the PQM is combined as $\mathbf{F}_\nu^\mathrm{QM}+\mathbf{F}_\nu^\mathrm{PQM} =  -\sum_{k,l}\mathbf{D}_{\nu,kl}\left(\sum_\mu\frac{\bm{\mathcal{P}}_\mu+\bm{\mathcal{Q}}_{\mu,kl}}{2M_\mu}\cdot\mathbf{D}_{\mu,kl}\right)|C_l|^2|C_k|^2$, the force from the divergence of $\mathbf{D}_{\nu,jk}$ is $\mathbf{F}^\mathrm{Div}_\nu = -\sum_{k,l}\mathbf{D}_{\nu,kl}\left(\sum_\mu\frac{\nabla_\mu\cdot\mathbf{D}_{\mu,kl}}{2M_\mu}\right)|C_l|^2|C_k|^2$, and the force from the phase correction is $\mathbf{F}^\mathrm{Ph}_\nu=\sum_{k,l}\frac{\nabla_\nu|C_k|}{|C_k|}\left(\sum_\mu\frac{|\mathbf{D}_{\mu,kl}|^2}{M_\mu}\right)|C_k|^2|C_l|^2$ and $\mathbf{F}^\mathrm{GI} = -\nabla_\nu\left(\sum_\mu\frac{\hbar^2\braket{\nabla_\mu\ewf}{\nabla_\mu\ewf}_e-|\bA_\mu|^2}{2M_\mu}\right)$ in Eq~(\ref{eq:force_exp}).

\subsubsection{Nuclear gradient of gauge independent part of time-dependent scalar potential}
To expand $\bF_\nu^\mathrm{GI}$ with the BO basis, we expand it up to the cartesian coordinates:
\begin{equation}
\begin{split}
    F^\mathrm{GI}_{\nu,q} =& -\pq\sum_{\mu,p}\dfrac{\hbar^2\braket{\pp\ewf}{\pp\ewf}_e-\hbar^2\left|\braket{\ewf}{\pp\ewf}\right|^2}{2M_\mu}\\
    =&-\pq\sum_{\mu,p}\dfrac{\hbar^2\sum_k\pp C_k^*\pp C_k-\hbar^2\left|\sum_kC_k^*\pp C_k\right|^2}{2M_\mu}\\
    =&-\sum_{\mu,p}\dfrac{\hbar^2}{2M_\mu}\left[\sum_k2\Re\big\{\pq\pp C_k^*\pp C_k\big\}-\sum_{k,l}2\Re\Big\{\big(\pq C^*_k\pp C_k+C_k^*\pq\pp C_k\big)\left(C_l\pp C_l^*\right)\Big\}\right].
\end{split}
\end{equation}
where $\{p, q\} \in \{x,y,z\}$, $\partial_{\nu,p}=\partial/\partial p_\nu$, and again the NAC terms are neglected.
The first and second order nuclear gradient of a BO coefficient are
\begin{equation}
    \pq C_k=\left(\dfrac{\pq|C_k|}{|C_k|}+\dfrac{i}{\hbar}\pq S_k\right)C_k,
\end{equation}
and
\begin{equation}
\begin{split}
    \pq\pp C_k=&\pq\left\{\left(\dfrac{\pp|C_k|}{|C_k|}+\dfrac{i}{\hbar}\pp S_k\right)C_k\right\}\\
        =&\left[\pq\left(\dfrac{\pp|C_k|}{|C_k|}+\dfrac{i}{\hbar}\pp S_k\right)\right]C_k+\left(\dfrac{\pp|C_k|}{|C_k|}+\dfrac{i}{\hbar}\pp S_k\right)\left(\dfrac{\pq|C_k|}{|C_k|}+\dfrac{i}{\hbar}\pq S_k\right)C_k,
\end{split}
\end{equation}
respectively, without NACs.
Then,
\begin{equation}
\begin{split}
    F^\mathrm{GI}_{\nu,q}
    =-\sum_{\mu,p}\dfrac{\hbar^2}{2M_\mu}\Bigg[\sum_k2\Re\bigg\{&
        \left[\pq\left(\dfrac{\pp|C_k|}{|C_k|}-\dfrac{i}{\hbar}\pp S_k\right)\right]\left(\dfrac{\pp|C_k|}{|C_k|}+\dfrac{i}{\hbar}\pp S_k\right)|C_k|^2\\
        &+\left(\dfrac{\pq|C_k|}{|C_k|}-\dfrac{i}{\hbar}\pq S_k\right)\left(\dfrac{\pp|C_k|}{|C_k|}-\dfrac{i}{\hbar}\pp S_k\right)\left(\dfrac{\pp|C_k|}{|C_k|}+\dfrac{i}{\hbar}\pp S_k\right)|C_k|^2\bigg\}\\
    -\sum_{k,l}2\Re\bigg\{&\left(\dfrac{\pq|C_k|}{|C_k|}-\dfrac{i}{\hbar}\pq S_k\right)\left(\dfrac{\pp|C_k|}{|C_k|}+\dfrac{i}{\hbar}\pp S_k\right)\left(\dfrac{\pp|C_l|}{|C_l|}-\dfrac{i}{\hbar}\pp S_l\right)    |C_k|^2|C_l|^2\\
        &+\left[\pq\left(\dfrac{\pp|C_k|}{|C_k|}+\dfrac{i}{\hbar}\pp S_k\right)\right]\left(\dfrac{\pp|C_l|}{|C_l|}-\dfrac{i}{\hbar}\pp S_l\right)|C_k|^2|C_l|^2\\
        &+\left(\dfrac{\pq|C_k|}{|C_k|}+\dfrac{i}{\hbar}\pq S_k\right)\left(\dfrac{\pp|C_k|}{|C_k|}+\dfrac{i}{\hbar}\pp S_k\right)\left(\dfrac{\pp|C_l|}{|C_l|}-\dfrac{i}{\hbar}\pp S_l\right)|C_k|^2|C_l|^2\bigg\}\Bigg]\\
    =-\sum_{\mu,p}\dfrac{\hbar^2}{2M_\mu}\Bigg[\sum_k2\Re\bigg\{&
        \left[\pq\left(\dfrac{\pp|C_k|}{|C_k|}-\dfrac{i}{\hbar}\pp S_k\right)\right]\left(\dfrac{\pp|C_k|}{|C_k|}+\dfrac{i}{\hbar}\pp S_k\right)|C_k|^2\\
        &+\left(\dfrac{\pq|C_k|}{|C_k|}-\dfrac{i}{\hbar}\pq S_k\right)\left(\left(\dfrac{\pp|C_k|}{|C_k|}\right)^2+\dfrac{1}{\hbar^2}(\pp S_k)^2\right)|C_k|^2\bigg\}\\
    -\sum_{k,l}2\Re\bigg\{&\left(\dfrac{\pq|C_k|}{|C_k|}-\dfrac{i}{\hbar}\pq S_k\right)\left(\dfrac{\pp|C_k|}{|C_k|}+\dfrac{i}{\hbar}\pp S_k\right)\left(-\dfrac{i}{\hbar}\pp S_l\right)    |C_k|^2|C_l|^2\\
        &+\left[\pq\left(\dfrac{\pp|C_k|}{|C_k|}+\dfrac{i}{\hbar}\pp S_k\right)\right]\left(-\dfrac{i}{\hbar}\pp S_l\right)|C_k|^2|C_l|^2\\
        &+\left(\dfrac{\pq|C_k|}{|C_k|}+\dfrac{i}{\hbar}\pq S_k\right)\left(\dfrac{\pp|C_k|}{|C_k|}+\dfrac{i}{\hbar}\pp S_k\right)\left(-\dfrac{i}{\hbar}\pp S_l\right)|C_k|^2|C_l|^2\bigg\}\Bigg],
\end{split}
\end{equation}
where terms that goes to zero when expanded and summed over $l$ are removed.

If we collect the real part, we get
\begin{equation}
\begin{split}
F^\mathrm{GI}_{\nu,q}
    =-\sum_{\mu,p}\dfrac{\hbar^2}{M_\mu}\Bigg[\sum_k\bigg\{&\pq\left(\dfrac{\pp|C_k|}{|C_k|}\right)\dfrac{\pp|C_k|}{|C_k|}+\dfrac{1}{\hbar^2}\pq\pp S_k\pp S_k\\
        &+\dfrac{\pq|C_k|}{|C_k|}\left(\dfrac{\pp|C_k|}{|C_k|}\right)^2+\dfrac{\pq|C_k|}{|C_k|}\dfrac{1}{\hbar^2}(\pp S_k)^2\bigg\}|C_k|^2\\
    -\sum_{k,l}&\bigg\{
        \dfrac{\pq|C_k|}{|C_k|}\dfrac{1}{\hbar^2}\pp S_k \pp S_l-\dfrac{1}{\hbar^2}\pq S_k\dfrac{\pp|C_k|}{|C_k|}\pp S_l\\
        &+\dfrac{1}{\hbar^2}\pq\pp S_k\pp S_l\\
        &+\dfrac{\pq|C_k|}{|C_k|}\dfrac{1}{\hbar^2}\pp S_k \pp S_l+\dfrac{1}{\hbar^2}\pq S_k\dfrac{\pp|C_k|}{|C_k|}\pp S_l\bigg\}|C_k|^2|C_l|^2\Bigg]\\
    =-\sum_{\mu,p}\dfrac{\hbar^2}{M_\mu}\Bigg[\sum_{k,l}\bigg\{&\pq\left(\dfrac{\pp|C_k|}{|C_k|}\right)\dfrac{\pp|C_k|}{|C_k|}+\dfrac{\pq|C_k|}{|C_k|}\left(\dfrac{\pp|C_k|}{|C_k|}\right)^2\\
        &+\dfrac{1}{\hbar^2}\pq\pp S_k\Big(\pp S_k-\pp S_l\Big)\\
        &+\dfrac{\pq|C_k|}{|C_k|}\dfrac{1}{\hbar^2}\pp S_k\Big(\pp S_k - 2 \pp S_l\Big)\bigg\}|C_k|^2|C_l|^2\Bigg]\\
    =-\sum_{\mu,p}\dfrac{\hbar^2}{M_\mu}\Bigg[\sum_{k,l}\bigg\{&\pq\left(\dfrac{\pp|C_k|}{|C_k|}\right)\dfrac{\pp|C_k|}{|C_k|}+\dfrac{\pq|C_k|}{|C_k|}\left(\dfrac{\pp|C_k|}{|C_k|}\right)^2\\
        &+\dfrac{1}{\hbar^2}\pq\pp S_k\Big(\pp S_k-\pp S_l\Big)\\
        &+\dfrac{\pq|C_k|}{|C_k|}\dfrac{1}{\hbar^2}\Big((\pp S_k)^2 - 2 \pp S_k\pp S_l\Big)\\
        &+\dfrac{\pq|C_k|}{|C_k|}\dfrac{1}{\hbar^2}(\pp S_l)^2\bigg\}|C_k|^2|C_l|^2\Bigg]\\
    =-\sum_{\mu,p}\dfrac{\hbar^2}{M_\mu}\Bigg[\sum_{k,l}\bigg\{&\pq\left(\dfrac{\pp|C_k|}{|C_k|}\right)\dfrac{\pp|C_k|}{|C_k|}+\dfrac{\pq|C_k|}{|C_k|}\left(\dfrac{\pp|C_k|}{|C_k|}\right)^2\\
        &+\dfrac{1}{\hbar^2}\pq\pp S_k\Big(\pp S_k-\pp S_l\Big)\\
        &+\dfrac{\pq|C_k|}{|C_k|}\dfrac{1}{\hbar^2}(\pp S_k-\pp S_l)^2\bigg\}|C_k|^2|C_l|^2\Bigg],
\end{split}
\end{equation}
where we added a term that goes to zero to symmetrize the equation.
Neglecting $O(\hbar^2)$ terms, we get
\begin{equation}
\begin{split}
F^\mathrm{GI}_{\nu,q}
    =-\sum_{\mu}\dfrac{\hbar^2}{M_\mu}\Bigg[\sum_{k,l}\bigg\{&\dfrac{1}{\hbar^2}\sum_p\pq\pp S_k\Big(\pp S_k-\pp S_l\Big)\\
        &+\dfrac{\pq|C_k|}{|C_k|}\dfrac{1}{\hbar^2}(\pp S_k-\pp S_l)^2\bigg\}|C_k|^2|C_l|^2\Bigg]\\
\end{split}
\end{equation}
As a result, we can write $\bF_\nu^\mathrm{GI}$ in terms of matrix multiplication:
\begin{equation}
\begin{split}
\bF^\mathrm{GI}_{\nu}
    =&-\sum_{k,l}\sum_{\mu}\left(\dfrac{\mathbf{H}_{\nu\mu,kl}\mathbf{D}_{\mu,kl}}{2M_\mu}\right)|C_k|^2|C_l|^2\\
    &-\sum_{k,l}\dfrac{\nabla_\nu|C_k|}{|C_k|}\left(\sum_{\mu}\dfrac{|\mathbf{D}_{\mu,kl}|^2}{M_\mu}\right)|C_k|^2|C_l|^2\Bigg],
\end{split}
\end{equation}
where $\mathbf{H}_{\nu\mu,kl}$ is a $3\times3$ cartesian sub-Hessian matrix of phase difference for nucleus $\nu$ and $\mu$, i.e. $\left[\mathbf{H}_{\nu\mu,kl}\right]_{pq} = \partial_{\nu p}\partial_{\mu q}(S_k-S_l)$.

Finally, if we substitute above for $\bF^\mathrm{GI}_\nu$ we get the nuclear force in terms of BO representation as:
\begin{equation}
\begin{split}
    \bF_\nu=&-\sum_k|C_k|^2\nabla_\nu\epsilon_k - \sum_{k,l}C_kC_l^*(\epsilon_k-\epsilon_l)\bd_{\nu,lk}\\
    &-\sum_{k,l}\mathbf{D}_{\nu,kl}\left\{\sum_\mu\dfrac{1}{2M_\mu}\left(
    \bm{\mathcal{P}}_{\mu}+\bm{\mathcal{Q}}_{\mu,kl}+\nabla_\mu\right)
    \cdot\mathbf{D}_{\mu,kl}\right\}|C_l|^2|C_k|^2\\
    &-\sum_{k,l}\sum_{\mu}\left(\dfrac{\mathbf{H}_{\nu\mu,kl}\mathbf{D}_{\mu,kl}}{2M_\mu}\right)|C_k|^2|C_l|^2,
\end{split}
\end{equation}
where $\bF_\nu^\mathrm{Ph}$ is canceled by a component of $\bF^{\mathrm{GI}}_\nu$.

\subsection{Numerical implementation}
We now introduce the numerical implementation of the new exact mixed quantum-classical equations derived in this work.
The reduced PQM $\bm{\mathcal{Q}}_{\nu,jk}$ can be calculated using the "bare" PQM: $\bm{\mathcal{G}}_{\nu,jk}=\nabla_\nu|\chi_j|^2/|\chi_j|^2+\nabla_\nu|\chi_k|^2/|\chi_k|^2=\bm{\mathcal{Q}}_{\nu,jk}+2\bm{\mathcal{P}}_\nu$, where $\chi_i=C_i\chi$ is the BO projected nuclear wave function.
Therefore, we first define the BO projected nuclear density for each state as a multi-dimensional Gaussian function based on the nuclear configurations and BO populations of trajectories~\cite{AgostiniJCTC2016} as
\begin{equation}
    |\chi_j(\dulR,t)|^2 \equiv \dfrac{\langle|C_j|^2\rangle(t)}{\mathcal{N}_j(t)} 
    \prod_\nu \exp\left(-\dfrac{|\bR_\nu-\bR_{\nu,j}(t)|^2}{2\sigma_{\nu,j}(t)^2}\right),
\end{equation}
where $\mathcal{N}_i$ is the normalization factor.
The Gaussian center and variance for each state are then defined as follows:
\begin{equation}
\begin{split}
    \bR_{\nu,j}(t) &\equiv \dfrac{\sum_{I}^{N_\mathrm{traj}} |C_{j}^{(I)}(t)|^2 \bR_\nu^{(I)}(t)}
    {N_\mathrm{traj}\,\langle|C_j|^2\rangle(t)} \\
    &= \dfrac{\sum_{I}^{N_\mathrm{traj}} |C_{j}^{(I)}(t)|^2 \bR_\nu^{(I)}(t)}
    {\sum_{J}^{N_\mathrm{traj}} |C_{j}^{(J)}(t)|^2},
\end{split}
\end{equation}
\begin{equation}
    \sigma_{\nu,j}(t)^2 \equiv
    \dfrac{\sum_{I}^{N_\mathrm{traj}} |C_{j}^{(I)}(t)|^2 
    \big|\bR_\nu^{(I)}(t)-\bR_{\nu,j}(t)\big|^2}
    {\sum_{J}^{N_\mathrm{traj}}|C_{j}^{(J)}(t)|^2},
\end{equation}
with $N_\mathrm{traj}$ denoting the number of trajectories.
Then, the total quantum momentum $\bm{\mathcal{P}}_\nu$ can be calculated from the total nuclear density, defined as the sum of the BO projected nuclear densities in the XF formalism: $|\chi(\dulR,t)|^2 = \sum_k |\chi_k(\dulR,t)|^2$.
As a result, we get
\begin{equation}
    \bm{\mathcal{G}}_{\nu,jk}^{(I)}=-\left[
        \left(\dfrac{1}{\sigma^2_{\nu,j}}+\dfrac{1}{\sigma^2_{\nu,k}}\right)\bR_\nu^{(I)}
        -\left(\dfrac{\bR_{\nu,j}}{\sigma^2_{\nu,j}}+\dfrac{\bR_{\nu,k}}{\sigma^2_{\nu,k}}\right)\right]
\end{equation}
and
\begin{equation}
    \bm{\mathcal{P}}_{\nu}^{(I)}=-\left[
        \sum_l\left(\dfrac{|C^{(I)}_l|^2}{\sigma^2_{\nu,l}}\right)\bR_\nu^{(I)}
        -\sum_l\left(\dfrac{|C^{(I)}_l|^2\bR_{\nu,l}}{\sigma^2_{\nu,l}}\right)\right].
\end{equation}

The second derivative terms $\mathbf{H}_{\nu\mu,kl}$ and $\nabla_\nu\cdot\mathbf{D}_{\nu,kl}$ can be approximated as time-integral of Hessian matrix elements of BO potential energy. However, it would make simulations more computationally expensive.
Therefore, we introduce $\Delta_{\nu,jk}$ to replace $\nabla_\nu\cdot \mathbf{D}_{\nu,jk}$ and ensure the averaged BO population conservation in the absence of NAC and the vanishing $\nabla_\nu S_j$ at $\dulR \rightarrow \infty$ condition for exact dynamics, and neglect $\mathbf{H}_{\nu\mu,kl}$.
From the averaged BO population conservation condition:
\begin{equation}
\begin{split}
    \sum_I\dfrac{d}{dt}|C^{(I)}_j|^2=-\sum_I\sum_\nu\dfrac{1}{M_\nu}\sum_k\left\{\left(
    \bm{\mathcal{Q}}^{(I)}_{\nu,jk}+\bm{\mathcal{P}}^{(I)}_{\nu,jk}\right)
    \cdot\mathbf{D}^{(I)}_{\nu,jk}+\Delta_{\nu, jk}\right\}|C^{(I)}_k|^2|C^{(I)}_j|^2 = 0,
\end{split}
\end{equation}
we define a uniform pair-wise shift term $\Delta_{\nu,ik}$ for each nucleus as
\begin{equation}
\begin{split}
    \Delta_{\nu,jk}=-\dfrac{\sum_I\left\{\left(
    \bm{\mathcal{Q}}^{(I)}_{\nu,jk}+\bm{\mathcal{P}}^{(I)}_{\nu,jk}\right)
    \cdot\mathbf{D}^{(I)}_{\nu,jk}\right\}|C^{(I)}_k|^2|C^{(I)}_j|^2}{\sum_J|C^{(J)}_k|^2|C^{(J)}_j|^2}.
\end{split}
\end{equation}

Meanwhile, $\nabla_\nu S^{(I)}_j$, which previously named as ``phase term'', can be approximated in different ways~\cite{MinPRL2015, AgostiniJCTC2016, HaJCP2022, ArribasJCP2023}.
In this work, we approximate $\nabla_\nu S^{(I)}_j$ to a BO momentum $\bP^{(I)}_{\nu,j}\equiv\alpha^{(I)}_j\bP^{(I)}_\nu$ which can be calculated based on the total energy conservation condition introduced in the main text where we used $E^{(I)}_\mathrm{tot}(t=0)$ in this work to use a constant value for stability.
Finally, dropping the terms merely contribute to the global phase, the working MQC equations for the new CTMQC method are obtained as follows:
\begin{equation}\label{eq:tdse_bo2}
\begin{split}
    \dot{C}^{(I)}_j=&-\dfrac{i}{\hbar}\epsilon^{(I)}_jC^{(I)}_j-\sum_k\sum_\nu\dfrac{\bP^{(I)}_\nu}{M_\nu}\cdot\bd^{(I)}_{\nu,jk}C^{(I)}_k\\
    &-\sum_\nu\sum_k\Bigg\{\dfrac{\bm{\mathcal{G}}^{(I)}_{\nu,jk}-\bm{\mathcal{P}}^{(I)}_\nu}{2M_\nu}
    \cdot\mathbf{D}^{(I)}_{\nu,jk}+\Delta_{\nu,jk}\Bigg\}|C^{(I)}_k|^2C^{(I)}_j\\
    &-\dfrac{i}{\hbar}\bigg(\sum_\nu\dfrac{|\bP^{(I)}_{\nu,j}|^2}{2M_\nu}-\dfrac{\bP^{(I)}_{\nu,j}\cdot\bP^{(I)}_\nu}{M_\nu}\bigg)C^{(I)}_j,
\end{split}
\end{equation}
and
\begin{equation}\label{eq:force_bo2}
\begin{split}
    \mathbf{F}_\nu^{(I)}=&-\sum_k|C^{(I)}_k|^2\nabla_\nu\epsilon^{(I)}_k 
    -\sum_{k,l}C^{(I)}_kC_l^{(I)*}(\epsilon_k-\epsilon_l)^{(I)}\bd^{(I)}_{\nu,lk}\\
    &-\sum_{k,l}\mathbf{D}^{(I)}_{\nu,kl}\left\{\sum_\mu\dfrac{\bm{\mathcal{G}}^{(I)}_{\mu,kl}-\bm{\mathcal{P}}^{(I)}_\mu}{2M_\mu}
    \cdot\mathbf{D}^{(I)}_{\mu,kl}+\Delta_{\nu,kl}\right\}|C^{(I)}_l|^2|C^{(I)}_k|^2.
\end{split}
\end{equation}
Compared to the original coupled trajectory MQC (CTMQC) equations, the total quantum momentum is changed to an "effective" quantum momentum $\bm{\mathcal{G}}_{\nu,jk}-\bm{\mathcal{P}}_\nu$ for each pair of states, and the phase correction term and $\Delta_{\nu,jk}$ term are added.

The new CTMQC simulations can be performed by following the steps below.
First, we select a set of nuclear configurations and momenta using an appropriate sampling method (e.g. Wigner or Boltzmann sampling), along with the initial BO state that each trajectory occupies.
For each trajectory, a static calculation of BO potential energies, their gradients, and NACs is performed at the corresponding nuclear position.
Then, we compute $\mathbf{D}_{\nu,lk}$ from the total energy and the current momentum of each trajectory.
Next, using the information from all trajectories, we calculate the QM and PQM for each trajectory, and calculate $\Delta_{\nu,lk}$.
We then evaluate the nuclear force and the time derivative of the BO coefficients according to Eqs.~(\ref{eq:force_bo2}) and (\ref{eq:tdse_bo2}), respectively, for each nuclear trajectory, so that we propagate the nuclear trajectory and corresponding electronic state.
This procedure is repeated from the static calculation step until the end of the simulation.

In the surface hopping based on exact factorization (SHXF) method, while the nuclear force and the hopping algorithm follow the conventional surface hopping (SH) scheme~\cite{TullyJCP1990}, PQM can be readily incorporated with negligible additional computational cost, as we can directly exploit the already-defined BO-projected nuclear density from the auxiliary trajectory on each state $\bR_{\nu,i}=\bR^{(I)}_{\nu,i}$~\cite{HaJPCL2018, HaJCP2022}.
The variance $\sigma^2_\nu$ is set identical for all states.
As a result,
\begin{equation}
    \bm{\mathcal{G}}_{\nu,ik}^{(I)}=-\dfrac{2\bR_\nu^{(I)}-\bR^{(I)}_{\nu,i}-\bR^{(I)}_{\nu,k}}{\sigma^2_\nu}
\end{equation}
and
\begin{equation}
    \bm{\mathcal{P}}_{\nu}^{(I)}=-\dfrac{\bR_\nu^{(I)}-\sum_k\left(|C^{(I)}_k|^2\bR^{(I)}_{\nu,k}\right)}{\sigma^2_\nu}.
\end{equation}
The phase correction term in the SHXF method becomes identical to the phase-corrected SH (PCSH) method~\cite{ShenviJCP2011} if we use $\nabla_\nu S_j\equiv\bP_j$.

\section{Computational detail}
The molecular Hamiltonian for two-state model systems can be expressed as
\begin{equation}
    \hat{H}(\dulR) = \hat{T}_\mathrm{n}(\dulR) + \hat{H}_{BO}(\dulR),
\end{equation}
where $\hat{T}_\mathrm{n}$ and $\hat{H}_\mathrm{BO}$ are the nuclear kinetic energy and the BO Hamiltonian operators, respectively.
The BO Hamiltonian matrix in the diabatic basis is
\begin{equation}\label{eq:H_BO}
    \hat{H}_\mathrm{BO} =
    \begin{pmatrix}
        V_{11}(\dulR) & V_{12}(\dulR) \\
        V_{21}(\dulR) & V_{22}(\dulR)
    \end{pmatrix},
\end{equation}
where $V_{ij}$ is defined for each model.

For dual avoided crossing (DAC), double arch geometry (DAG), and two-dimensional non-separable (2DNS) models, the exact quantum wave-packet dynamics is performed using the split-operator technique~\cite{FeitJCP1982} on a spatial grid for the nuclear degrees of freedom, with a time step of 0.1 a.u.
A time step of 1.0 a.u. is used for the MQC simulations in both models.
The mass of the nuclear degrees of freedom is set to $2000.0$ a.u.
For SH-based dynamics simulations, we use 1000 trajectories, whereas the CT methods propagate 200 trajectories.

The standard deviation parameter for SHXF simulations is set to the same value as that of the initial nuclear density while the EDC and EDC2 parameters are taken from Refs.~\citenum{GranucciJCP2007} and~\citenum{XuJCP2019}, respectively.

The energy-based correction (EDC) scheme~\cite{GranucciJCP2007,ShaoJPCL2023} for SH methods rescales the BO coefficients of each trajectory with the decoherence time constant $\tau_{ja}$ at every time step as
\begin{align}
        C'_j = C_j\exp(-\Delta t/\tau_{ja})\\
        C'_a = C_a \sqrt{\dfrac{1-\sum_{j\neq a}|C'_j|^2}{|C_a|^2}},
\end{align}
where $a$ is the active state of the trajectory, so that the electronic state of the trajectory decoheres toward the active state.

The time constant for PCSH-EDC is defined~\cite{GranucciJCP2007} as
\begin{equation}
    \tau_{ja} = \dfrac{1}{|\epsilon_j-\epsilon_a|}\left(1+\dfrac{C}{E_\mathrm{kim}}\right),
\end{equation}
where $C=0.1$ a.u. and $E_\mathrm{kin}$ is the kinetic energy of a trajectory, while PCSH-EDC2 employs
\begin{equation}
    \tau_{ja} = C_0+C1\dfrac{\max(E_\mathrm{kin},\epsilon_j-\epsilon_a)}{E_\mathrm{kin}-(\epsilon_j-\epsilon_a)},
\end{equation}
where $C_0=2.5\times 10^{-3}$ and $C_1=2.0\time10^5$, which was proposed by Shao and coworkers in Ref.~\citenum{ShaoJPCL2023}.

\subsection{Double arch geometry model}
The diabatic Hamiltonian matrix elements of the DAG model are given by
\begin{align}
    &V_{11}(R) = a,\\
    &V_{22}(R) = -a,\\
    &V_{12}(R) = V_{21}(R)\nonumber\\
    &=
    \begin{cases}
        -b\exp(c(R-d))+b\exp(c(R+d))&\text{if } R<-d,\\
        -b\exp(c(R-d))-b\exp(-c(R+d))+2b&\text{if } -d\leq R\leq d,\\
        b\exp(-c(R-d))-b\exp(-c(R+d))&\text{if } R>d
    \end{cases},
\end{align}
where $a=6 \times 10^{-4}$, $b=0.1$, $c=0.9$, and $d=4.0$.

The initial molecular wave function is set to a nuclear Gaussian function centered at $R_0=-25.0$ a.u. on the lower BO state with a momentum delta-kick $k_0$ with the standard deviation $\sigma=2.0$ a.u., i.e.
\begin{equation}
    \ket{\Psi(R,t_0)} = \mathcal{N}\exp\left(-\dfrac{(R-R_0)^2}{4\sigma^2}\right)
        \exp\left(ik_0(R-R_0)\right)\ket{\phi_1}.
\end{equation}
The initial conditions of the nuclear trajectories are obtained from the Wigner sampling of the above wave function, and the initial BO coefficient for the lower state is set to 1.0.

\subsection{Dual avoided crossing}
The dual avoided crossing (DAC) model has the diabatic Hamiltonian:
\begin{equation}
    \begin{split}
        &V_{11}(R) = 0\\
        &V_{22}(R) = -a\exp(-bR^2) + e\\
        &V_{12}(R)=H_{21}(R)=c\exp(-dR^2)
    \end{split}
\end{equation}
with $a=0.1$, $b=0.28$, $c=0.015$, $d=0.06$, and $e=0.05$.

The initial conditions for the quantum dynamics and MQC dynamics simulations are generated in the same way as those used for the DAG model. The center of the Gaussian function is set to $R=-8.0$ a.u. with $\sigma=10.0 / k_0$ on the lower BO state.

\subsection{Two-dimensional non-separable model} 
The diabatic Hamiltonian matrix elements of the 2DNS model are given by
\begin{align}
    &V_{11}(x,y) = -E_0,\\
    &V_{22}(x,y) = -A\exp\left(-B(0.75(x+y)^2+0.25(x-y)^2)\right),\\
    &V_{12}(x,y) = V_{21}(x,y) = C\exp\left(-D(0.25(x+y)^2+0.75(x-y)^2)\right),
\end{align}
where $A=0.15, B=0.14, C=0.015, D=0.06,$ and $E_0=0.05$ a.u.

Similar to the DAG model simulations, the initial molecular wave function is set to a two-dimensional nuclear Gaussian function centered at $(x_0,y_0)=(-8.0, 0.0)$ a.u. on the lower BO state with a momentum delta-kick $(k_{0x}, k_{0y})=(k_0, 0.0)$ a.u. with the standard deviation $(\sigma_x, \sigma_y)=(0.5, 0.5)$ a.u., i.e.
\begin{equation}
    \ket{\Psi(x,y,t_0)} = \mathcal{N}\exp\left(-\dfrac{(x-x_0)^2}{4\sigma_x^2}\right)\exp\left(-\dfrac{y^2}{4\sigma_y^2}\right)
        \exp\left(ik_0(x-x_0)\right)\ket{\phi_1},
\end{equation}
and the trajectories are generated correspondingly with the Wigner sampling.
\clearpage
\section{Supplementary results}
\subsubsection{Transmission probability $T_1$}
\begin{figure}[h!]
    \centering
    \includegraphics[width=0.5\linewidth]{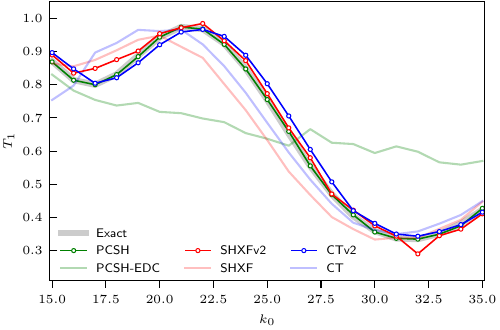}
    \caption{Transmission probability for the lower state ($T_1$) of the DAC model as a function of initial momentum $k_0$ for various dynamics methods.}
    \label{fig:RT_dac}
\end{figure}
\begin{figure}[h!]
    \centering
    \includegraphics[width=0.5\linewidth]{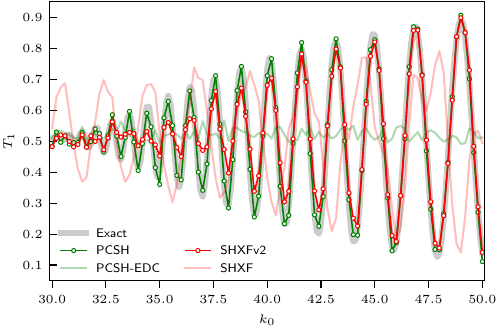}
    \caption{Transmission probability for the lower state ($T_1$) of the DAG model as a function of initial momentum $k_0$ for surface hopping-based methods obtained from the active states ($a^{(I)}(t)$) of trajectories: $T_1 = \sum_{I \in \{R^{(I)}(t_f)>0\}} \sum_I \delta_{1a^{(I)}(t_f)}/N_\mathrm{traj}$.}
    \label{fig:RT_run}
\end{figure}
\clearpage
\begin{figure}[h!]
    \centering
    \includegraphics[width=0.5\linewidth]{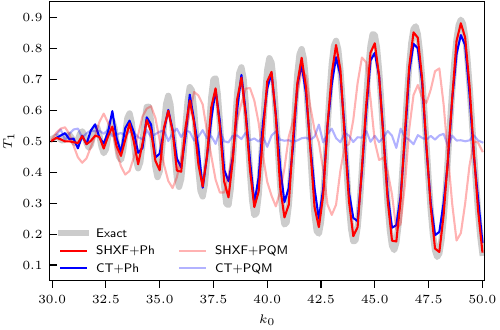}
    \caption{Transmission probability for the lower state ($T_1$) of the DAG model as a function of initial momentum $k_0$, calculated using the SHXF and CT methods with the PQM and phase terms included independently.}
    \label{fig:RT_pqm_ph}
\end{figure}

\subsubsection{Born-Oppenheimer population and coherence}
\begin{figure}[h!]
    \centering
    \includegraphics[width=0.5\linewidth]{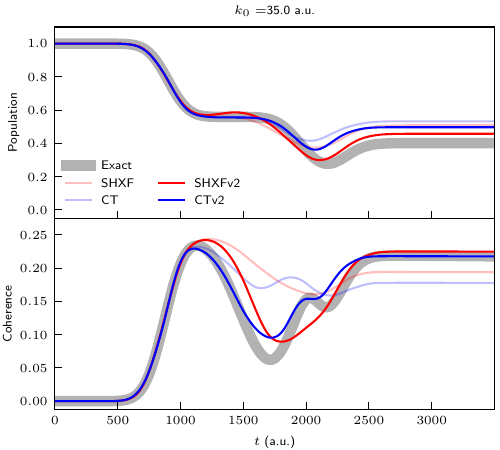}
    \caption{Time evolution of BO population $\langle|C_1|^2\rangle(t)$ (upper) and coherence $\langle|C_1C_2|^2\rangle(t)$ (lower) with initial momentum $k_0 = 35$ a.u. for the DAG model.}
    \label{fig:pop_coh_35}
\end{figure}

\begin{figure*}[h!]
    \centering
    \includegraphics[width=0.49\linewidth]{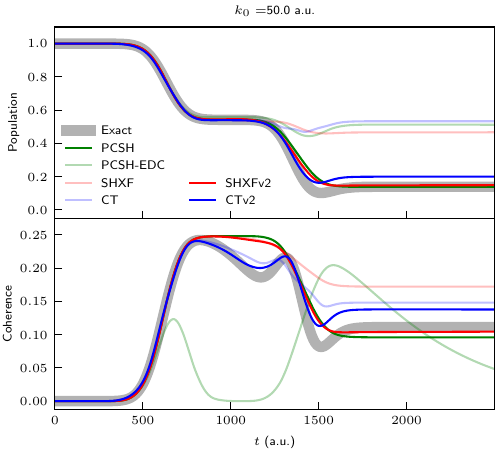}
    \caption{Time evolution of BO population $\langle|C_1|^2\rangle(t)$ (upper) and coherence $\langle|C_1C_2|^2\rangle(t)$ (lower) with initial momentum $k_0 = 50$ a.u. for the DAG model}
    \label{fig:pop_coh_50}
\end{figure*}

\begin{figure}
    \centering
    \includegraphics[width=0.5\linewidth]{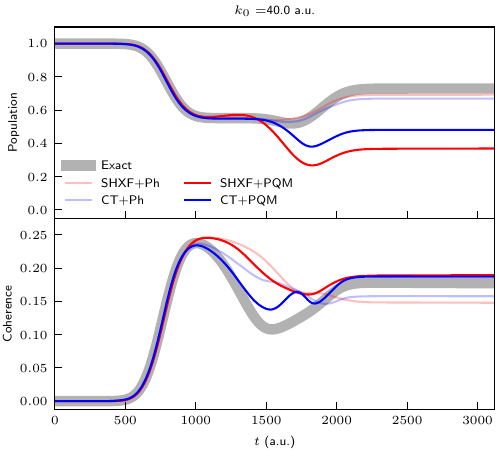}
    \caption{Time evolution of BO population $\langle|C_1|^2\rangle(t)$ (upper) and coherence $\langle|C_1C_2|^2\rangle(t)$ (lower) calculated using the SHXF and CT methods with the PQM and phase terms included independently with initial momentum $k_0 = 40$ a.u. for the DAG model.}
    \label{fig:pop_coh_pqm_ph}
\end{figure}

\begin{figure}[h!]
    \centering
    \includegraphics[width=1.0\linewidth]{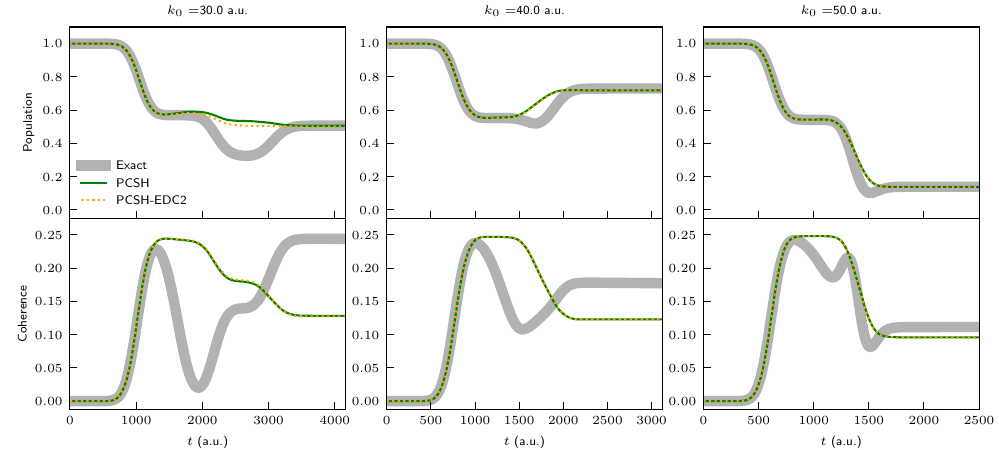}
    \caption{Time evolution of BO population $\langle|C_1|^2\rangle(t)$ (upper) and coherence $\langle|C_1C_2|^2\rangle(t)$ (lower) with different initial momenta $k_0 = 30, 40$ and $50$ a.u. obtained from PCSH-EDC2 for the DAG model.}
    \label{fig:pop_coh_edc2}
\end{figure}
\clearpage
\begin{figure}[h!]
    \centering
    \includegraphics[width=1.0\linewidth]{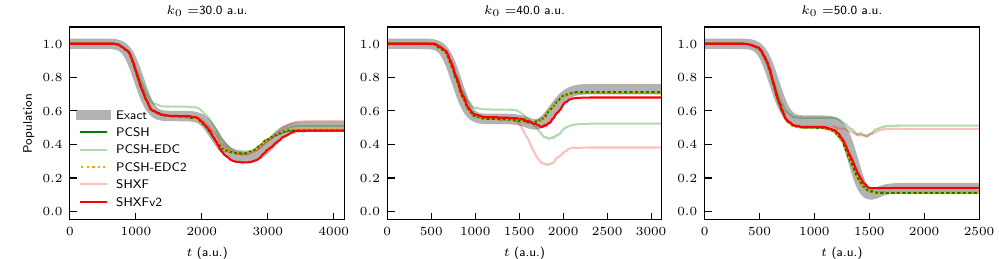}
    \caption{Time evolution of BO population calculated based on the active state $a^{(I)}(t)$ of trajectories: $P_i(t)=\sum_I \delta_{ia^{(I)}(t)}/N_\mathrm{traj}$, with different initial momenta $k_0 = 30, 40$ and $50$ a.u. obtained from the SH-based methods for the DAG model.}
    \label{fig:pop_run}
\end{figure}

\begin{figure}[h!]
    \centering
    \includegraphics[width=1.0\linewidth]{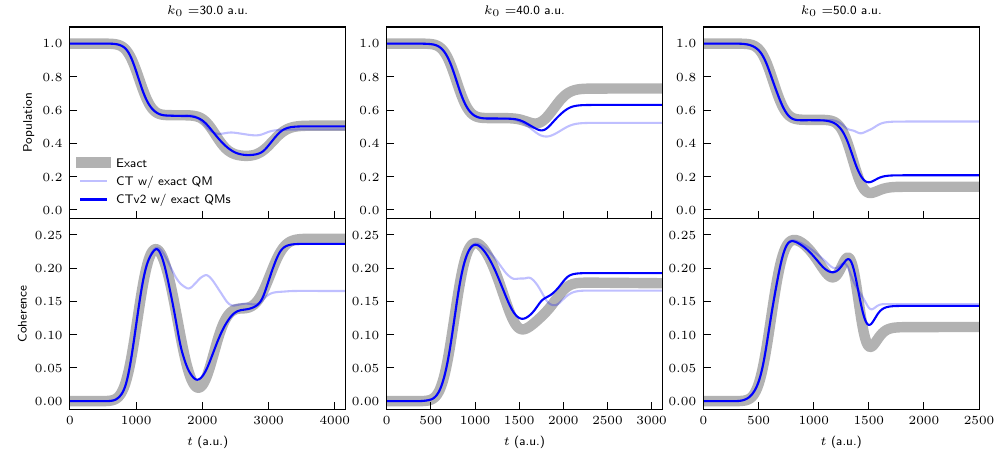}
    \caption{Time evolution of BO population $\langle|C_1|^2\rangle(t)$ (upper) and coherence $\langle|C_1C_2|^2\rangle(t)$ (lower) with different initial momenta $k_0 = 30, 40$ and $50$ a.u. obtained from the CT methods with the QM $\mathbf{\mathcal{P}}_\nu$ (and PQM, $\mathbf{\mathcal{Q}}_{\nu,ik}$ for CTv2) obtained from quantum dynamics for the DAG model.}
    \label{fig:pop_coh_qd}
\end{figure}
\clearpage

\subsubsection{Spatial distribution}
    \begin{figure}[h!]
        \centering
        \includegraphics[width=0.5\linewidth]{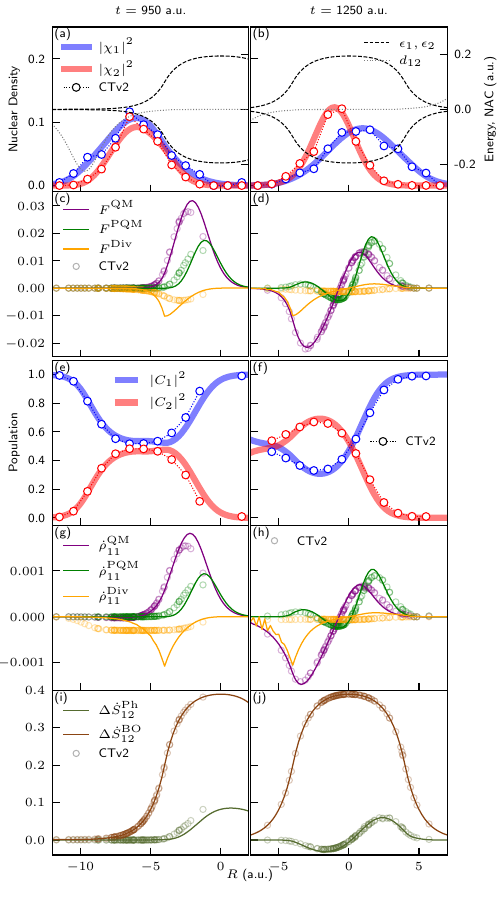}
        \caption{Spatial distribution of BO projected nuclear wave packets (a,b), individual components of the nuclear force (c,d), BO populations (e,f), and individual components of the time derivative of BO population for state 1 (g,h) and the phase difference (i,j) obtained from CTv2 in comparison with the values calculated from the exact quantum wave packet dynamics for $k_0=40.0$~a.u. The BO potential energies $\epsilon_{1/2}$ and the NAC $d_{12}$ are depicted to interpret wave packet dynamics (a,b).}
        \label{fig:element_spatial}
    \end{figure}

\begin{figure}[h!]
    \centering
    \includegraphics[width=0.49\linewidth]{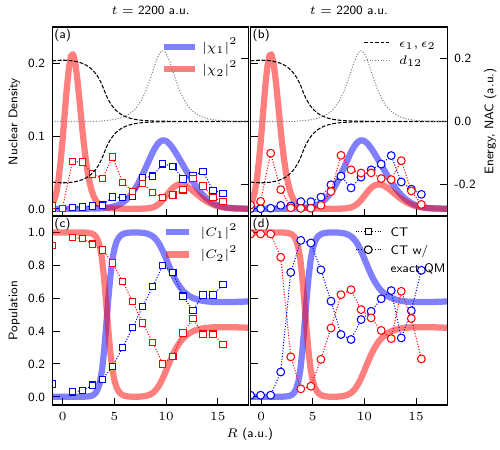}
    \caption{Spatial distribution of nuclear wave packets (a,b) and BO populations (c,d) at $t=2200$ a.u. with the initial momentum $k_0=30.0$ a.u. for CT (a,c) and CT with the QM, $\mathbf{\mathcal{P}}_\nu$, obtained from quantum dynamics (b,d). The BO potential energies $\epsilon_{1/2}$ and the NAC $d_{12}$ are depicted to interpret wave packet dynamics (a,b).}
    \label{fig:ct_spatial_k0_30}
\end{figure}

\begin{figure}[h!]
    \centering
    \includegraphics[width=0.49\linewidth]{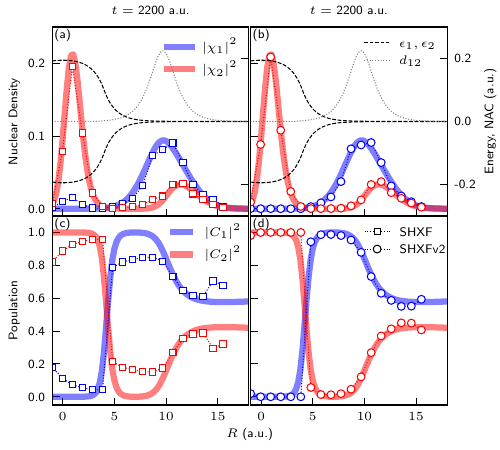}
    \caption{Spatial distribution of nuclear wave packets (a,b) and BO populations (c,d) at $t=2200$ a.u. with the initial momentum $k_0=30.0$ a.u. for SHXF (a,c) and SHXF2 (b,d) methods. BO potentials and NAC are depicted as in Fig.~\ref{fig:ct_spatial_k0_30}.}
    \label{fig:shxf_both_spatial_k0_30}
\end{figure}

\begin{figure}[h!]
    \centering
    \includegraphics[width=0.49\linewidth]{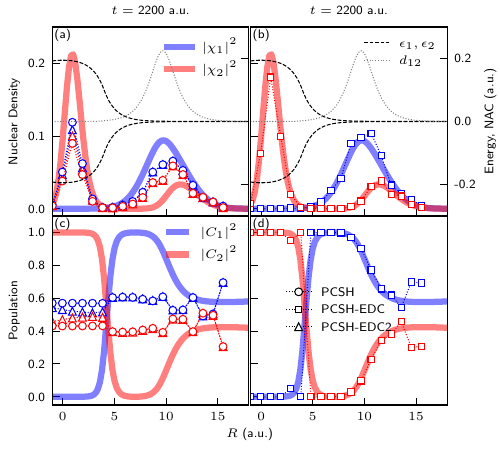}
    \caption{Spatial distribution of nuclear wave packets (a,b) and BO populations (c,d) at $t=2200$ a.u. with the initial momentum $k_0=30.0$ a.u. for PCSH, PCSH-EDC2 (a,c), and PCSH-EDC (b,d). BO potentials and NAC are depicted as in Fig.~\ref{fig:ct_spatial_k0_30}.}
    \label{fig:pcsh_both_spatial_k0_30}
\end{figure}

\begin{figure}[h!]
    \begin{subfigure}{0.49\textwidth}
        \includegraphics[width=1.0\textwidth]{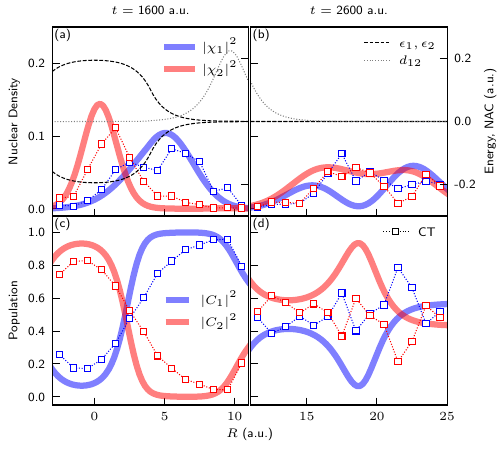}
    \end{subfigure}
    \begin{subfigure}{0.49\textwidth}
        \includegraphics[width=1.0\textwidth]{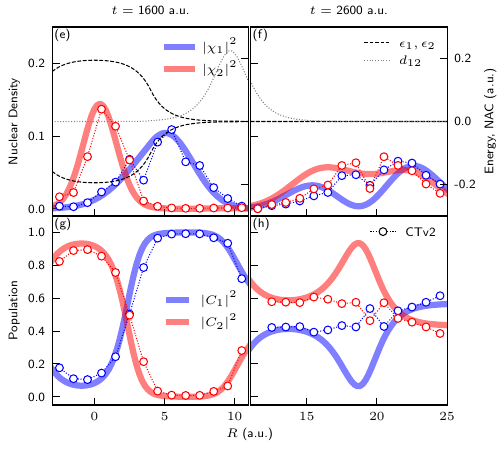}
    \end{subfigure}
    \caption{Spatial distribution of nuclear wave packets and BO populations at different time steps, $t=1600$ a.u. (a,c,e,g) and $2600$ a.u. (b,d,f,h) with the initial momentum $k_0=35.0$ a.u. for CT (a-d) and CTv2 (e-h). The BO potential energies $\epsilon_{1/2}$ and the NAC $d_{12}$ are depicted to interpret wave packet dynamics (a,b,e,f).}
    \label{fig:ct2_spatial_k0_35}
\end{figure}

\begin{figure}[h!]
    \begin{subfigure}{0.49\textwidth}
        \includegraphics[width=1.0\textwidth]{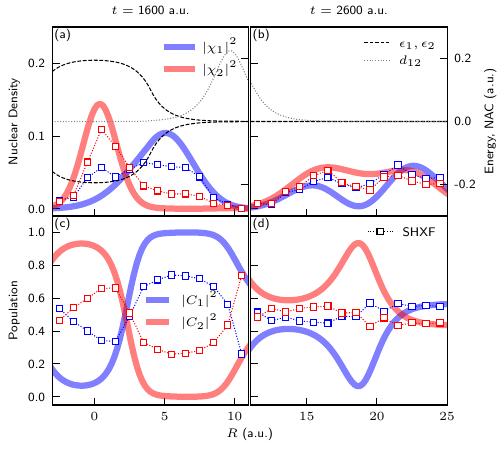}
    \end{subfigure}
    \begin{subfigure}{0.49\textwidth}
        \includegraphics[width=1.0\textwidth]{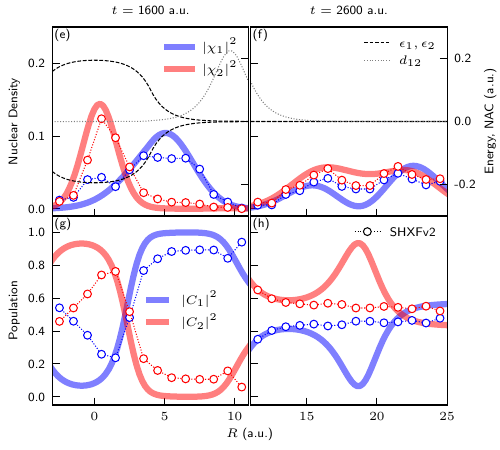}
    \end{subfigure}
    \caption{Spatial distribution of nuclear wave packets and BO populations at different time steps, $t=1600$ a.u. (a,c,e,g) and $2600$ a.u. (b,d,f,h) with the initial momentum $k_0=35.0$ a.u. for SHXF (a-d) and SHXFv2 (e-h). BO potentials and NAC are depicted as in Fig.~\ref{fig:ct2_spatial_k0_35}.}
    \label{fig:shxf2_spatial_k0_35}
\end{figure}

\begin{figure}[h!]
    \begin{subfigure}{0.49\textwidth}
        \includegraphics[width=1.0\textwidth]{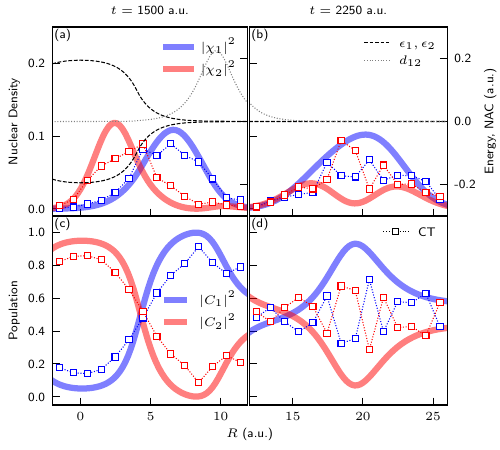}
    \end{subfigure}
    \begin{subfigure}{0.49\textwidth}
        \includegraphics[width=1.0\textwidth]{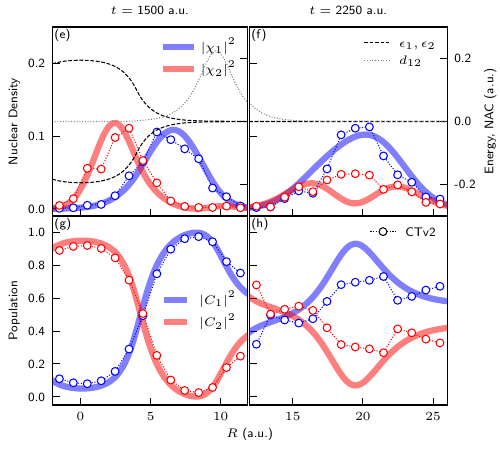}
    \end{subfigure}
    \caption{Spatial distribution of nuclear wave packets and BO populations at different time steps, $t=1500$ a.u. (a,c,e,g) and $2250$ a.u. (b,d,f,h) with the initial momentum $k_0=40.0$ a.u. for CT (a-d) and CT2 (e-h). BO potentials and NAC are depicted as in Fig.~\ref{fig:ct2_spatial_k0_35}.}
    \label{fig:ct_both_spatial_k0_40}
\end{figure}

\begin{figure}[h!]
    \begin{subfigure}{0.49\textwidth}
        \includegraphics[width=1.0\textwidth]{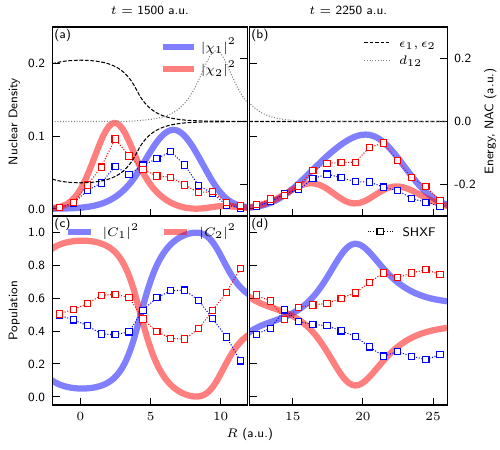}
    \end{subfigure}
    \begin{subfigure}{0.49\textwidth}
        \includegraphics[width=1.0\textwidth]{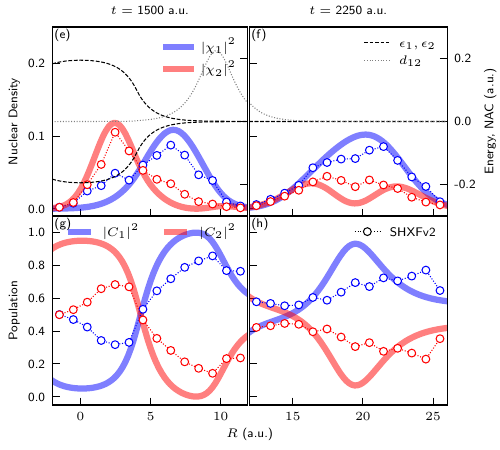}
    \end{subfigure}
    \caption{Spatial distribution of nuclear wave packets and BO populations at different time steps, $t=1500$ a.u. (a,c,e,g) and $2250$ a.u. (b,d,f,h) with the initial momentum $k_0=40.0$ a.u. for SHXF (a-d) and SHXF2 (e-h). BO potentials and NAC are depicted as in Fig.~\ref{fig:ct2_spatial_k0_35}.}
    \label{fig:shxf_both_spatial_k0_40}
\end{figure}

\begin{figure}[h!]
    \begin{subfigure}{0.49\textwidth}
        \includegraphics[width=1.0\textwidth]{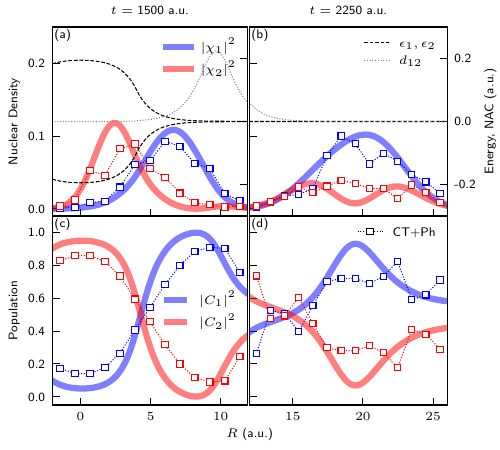}
    \end{subfigure}
    \begin{subfigure}{0.49\textwidth}
        \includegraphics[width=1.0\textwidth]{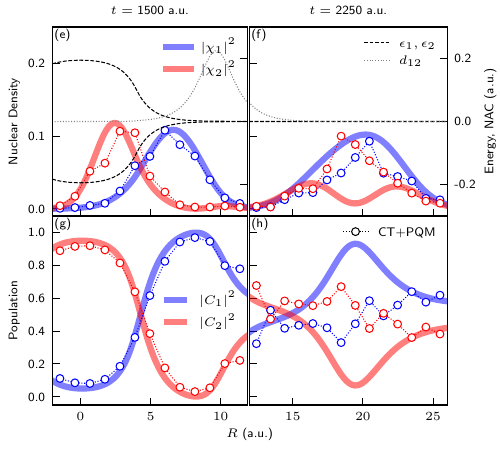}
    \end{subfigure}
    \caption{Spatial distribution of nuclear wave packets and BO populations at different time steps, $t=1500$ a.u. (a,c,e,g) and $2250$ a.u. (b,d,f,h) with the initial momentum $k_0=40.0$ a.u. for CT with the phase term (a-d) and CT with the PQM term (e-h). BO potentials and NAC are depicted as in Fig.~\ref{fig:ct2_spatial_k0_35}.}
    \label{fig:ct_pqm_ph_spatial_k0_40}
\end{figure}

\begin{figure}[h!]
    \begin{subfigure}{0.49\textwidth}
        \includegraphics[width=1.0\textwidth]{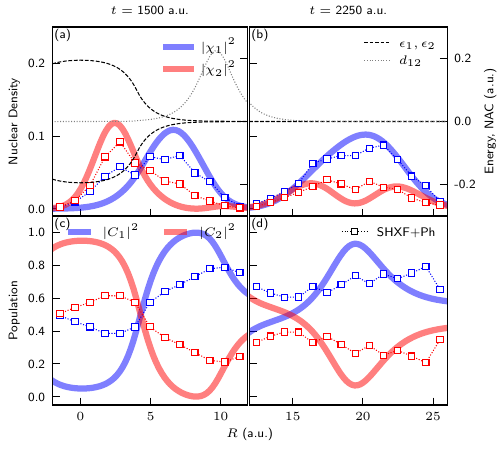}
    \end{subfigure}
    \begin{subfigure}{0.49\textwidth}
        \includegraphics[width=1.0\textwidth]{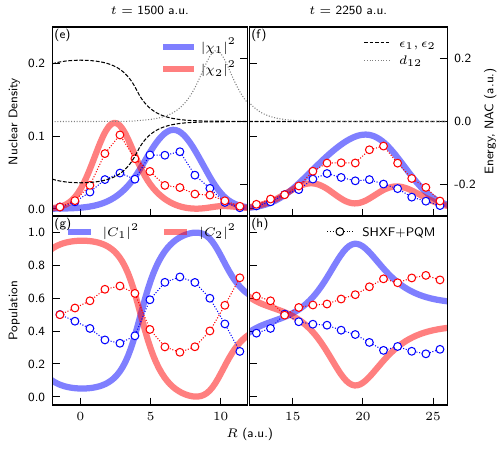}
    \end{subfigure}
    \caption{Spatial distribution of nuclear wave packets and BO populations at different time steps, $t=1500$ a.u. (a,c,e,g) and $2250$ a.u. (b,d,f,h) with the initial momentum $k_0=40.0$ a.u. for SHXF with the phase term (a-d) and SHXF with the PQM term (e-h). BO potentials and NAC are depicted as in Fig.~\ref{fig:ct2_spatial_k0_35}.}
    \label{fig:shxf_pqm_ph_spatial_k0_40}
\end{figure}

\begin{figure}[h!]
    \begin{subfigure}{0.49\textwidth}
        \includegraphics[width=1.0\textwidth]{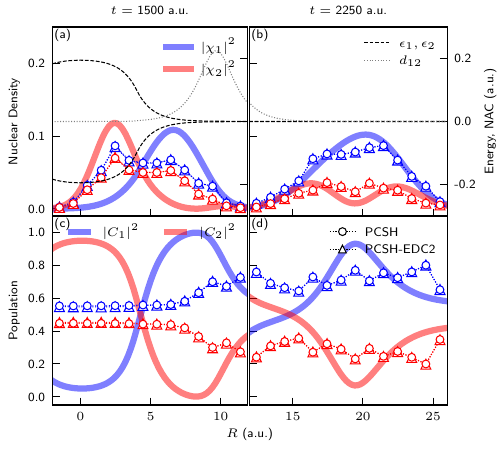}
    \end{subfigure}
    \begin{subfigure}{0.49\textwidth}
        \includegraphics[width=1.0\textwidth]{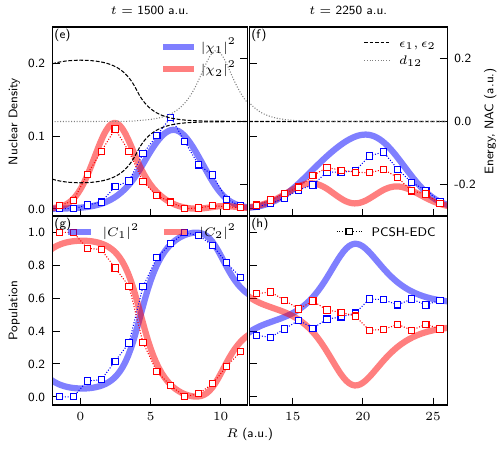}
    \end{subfigure}
    \caption{Spatial distribution of nuclear wave packets and BO populations at different time steps, $t=1500$ a.u. (a,c,e,g) and $2250$ a.u. (b,d,f,h) with the initial momentum $k_0=40.0$ a.u. for PCSH, PCSH-EDC2 (a-d), and PCSH-EDC (e-h). BO potentials and NAC are depicted as in Fig.~\ref{fig:ct2_spatial_k0_35}.}
    \label{fig:pcsh_both_spatial_k0_40}
\end{figure}

\begin{figure}[h!]
    \begin{subfigure}{0.49\textwidth}
        \includegraphics[width=1.0\textwidth]{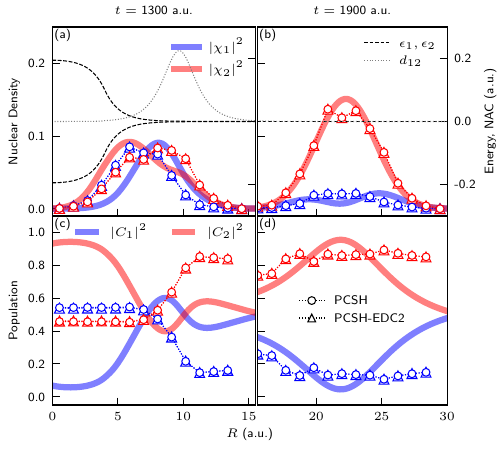}
    \end{subfigure}
    \begin{subfigure}{0.49\textwidth}
        \includegraphics[width=1.0\textwidth]{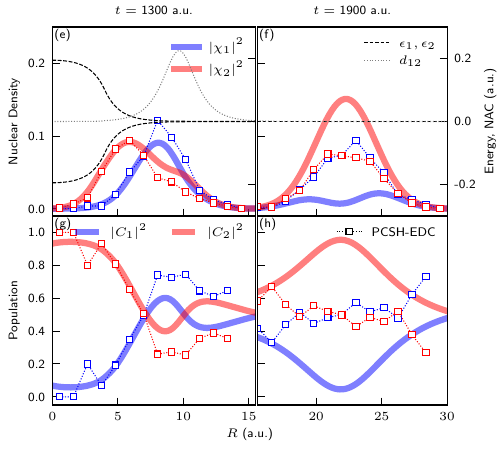}
    \end{subfigure}
    \caption{Spatial distribution of nuclear wave packets and BO populations at different time steps, $t=1300$ a.u. (a,c,e,g) and $1900$ a.u. (b,d,f,h) with the initial momentum $k_0=50.0$ a.u. for PCSH, PCSH-EDC2 (a-d), and PCSH-EDC (e-h). BO potentials and NAC are depicted as in Fig.~\ref{fig:ct2_spatial_k0_35}.}
    \label{fig:pcsh_both_spatial_k0_50}
\end{figure}

\clearpage
\subsubsection{Numerical stability}
\begin{figure}[h!]
    \centering
    \includegraphics[width=0.5\linewidth]{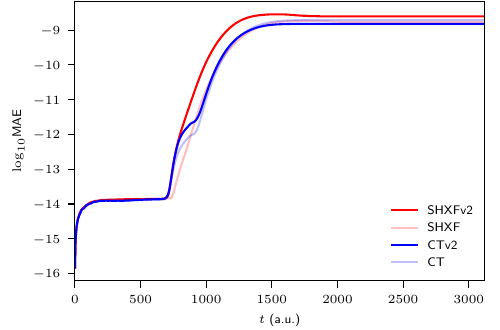}
    \caption{Mean absolute error (MAE) of the norm of the electronic state as a function of time for $k_0=40$ of the DAG model.}
    \label{fig:mae}
\end{figure}
\bibliography{references}